\documentclass[lettersize,journal]{IEEEtran}
\usepackage{amsmath,amsfonts}
\usepackage{algorithmic}
\usepackage{algorithm}
\usepackage{array}
\usepackage[caption=false,font=normalsize,labelfont=sf,textfont=sf]{subfig}
\usepackage{textcomp}
\usepackage{amsmath}
\usepackage{stfloats}
\usepackage{url}
\usepackage{booktabs}
\usepackage{verbatim}
\usepackage{graphicx}
\usepackage{cite}
\usepackage{listings}
\usepackage{amsmath,amsfonts,color}
\usepackage{graphicx}
\usepackage{amsfonts}
\usepackage{array}
\usepackage{amssymb}
\usepackage{helvet}
\usepackage{color}
\usepackage{bm}
\usepackage{float}
\usepackage{cite}
\usepackage{multirow}
\usepackage{amssymb}
\usepackage[caption=false,font=normalsize,labelfont=sf,textfont=sf]{subfig}
\usepackage{amsmath,amsfonts}
\usepackage{algorithmic}
\usepackage{algorithm}
\usepackage{textcomp}
\usepackage{stfloats}
\usepackage{url}
\usepackage{verbatim}
\usepackage{graphicx}
\usepackage{cite}
\usepackage{amssymb}
\usepackage{makecell}
\usepackage{mathrsfs}

\newcommand{\RNum}[1]{\uppercase\expandafter{\romannumeral #1\relax}}

\begin{document}

\bstctlcite{BSTcontrol}

\title{ISAC and Vision Fusion for Fine-Grained Low-Altitude Target Recognition}
\author{Zhonghua Chu, Hongliang Luo, Boxuan Sun, Shengjie Quan, Chuanbin Zhao, and Feifei Gao,~\IEEEmembership{Fellow,~IEEE}
\thanks{Z. Chu, H. Luo, B. Sun, S. Quan, C. Zhao and F. Gao are with Department of Automation, Tsinghua University, Beijing 100084, China (email: zhu-zh25@mails.tsinghua.edu.cn; luohl23@mails.tsinghua.edu.cn; sunbx26@163.com; quansj23@mails.tsinghua.edu.cn; zcb23@mails.tsinghua.edu.cn; feifeigao@ieee.org).
}
}



\maketitle

\begin{abstract}
In this paper, we propose an integrated sensing and communications (ISAC) and vision fusion framework for fine-grained  low-altitude target recognition.
Specifically, we first utilize  ISAC system to estimate the position of the  low-altitude target. 
Then we adjust the working parameters of the Pan-Tilt-Zoom (PTZ) camera based on the estimated  target position, such that the camera can capture the image of tiny low-altitude target from several hundred meters away.
After obtaining the wireless echo signal and visual image of low-altitude target, 
we employ the short-time Fourier transform (STFT) to obtain the micro-Doppler (mD) spectrum of the target from wireless echo signal, and design a conditional generative adversarial network (cGAN)-based denoising network to optimize the quality of the mD spectrum.
Meanwhile, we employ YOLOv11 to detect the low-altitude target from visual image, and then  crop the small-sized feature image of the target from the original image.
Next,  
we  design a fine-grained low-altitude target recognition network with MobileViT, which can fuse the optimized mD spectrum and the cropped feature image to distinguish the subcategory of low-altitude target.
Moreover, 
we generate a joint ISAC and vision dataset (JIVD) for low-altitude target monitoring based on AirSim and Wireless InSite, which includes diverse target subcategories, scenarios, and weather conditions.
The effectiveness and superiority of the  proposed    scheme   have been demonstrated by simulation results.

\end{abstract}


\section{Introduction}
Recently, low-altitude economy (LAE) has been experiencing  explosive development, which fosters a lot of novel industry applications, such as low-altitude logistics, low-altitude agriculture, low-altitude tourism, low-altitude  firefighting \cite{Jiang20236GNN, 10723207, 9275613,9666755, 10977743}.
The primary low-altitude targets within the airspace encompass unmanned aerial vehicles (UAVs), electric vertical take-off and landing (eVTOL) aircraft, helicopters, traditional fixed-wing aircraft, and naturally occurring birds 
\cite{11159257}. 
As the density of low-altitude targets surges, severe security threats such as unauthorized intrusions and potential collisions with wildlife inevitably arise \cite{9681624}.
Hence, industry and academia are committed to constructing the low-altitude  monitoring network to ensure airspace safety. 
In the low-altitude  monitoring network,
an important observation   is that different subcategories of low-altitude target possess  distinct physical characteristics and represent different levels of threats.
Therefore, it is necessary to build a fine-grained low-altitude target recognition system to distinguish the subcategories of the targets.

Conventional low-altitude target recognition systems primarily rely on radar and visual sensors \cite{10535988, 10569843, doi:10.1177/1729881420962907, 8353365}. 
On the one hand, 
radar-based systems  extract the unique micro-motion features and geometrical characteristics of the low-altitude target from radar echo signals, and then utilize machine learning or deep learning  to  distinguish the low-altitude targets.
For example,
S.~Li~\emph{et~al.} capture the subtle micro-Doppler (mD) signatures and high-resolution range profiles of UAVs, and design a Mamba-based network to recognize different subcategories of UAVs
\cite{11106829}.
B.-S.~Oh~\emph{et~al.} extract the statistical and geometrical features of UAVs by  empirical-mode decomposition (EMD), and  these features are subsequently   fed into a nonlinear support vector machine (SVM) for efficient UAV classification\cite{8239598}.
B.~Torvik~\emph{et~al.} extract the polarimetric intercorrelation and  decomposition features  of UAVs and birds, and then employ a nearest-neighbor classifier to realize  binary target recognition \cite{7508914}.  
On the other hand, visual-based systems  capture the  rich textural details, color information, and fine-grained semantic features of low-altitude targets from the visual image, thereby realizing  target recognition. 
For example, \textcolor{black}{ B.~Wang~\emph{et~al.} combine MobileNetV3 and coordinate attention with YOLOv5s \cite{khanam2024yolov5} to strengthen the spatial and textural information of low-altitude targets \cite{11083683}.}
D.~Dosi~\emph{et~al.} utilize YOLOv8 \cite{sohan2024review} to identify thermal contrast and morphological patterns of UAVs, which enables robust recognition and tracking in complex environments \cite{11213400}. 
 F.~Mahdavi~\emph{et~al.} utilize fisheye cameras to distinguish UAVs from birds by comparing deep-learned semantic features with traditional histogram of oriented gradients  descriptors \cite{9349620}.

Nonetheless,
due to the limited target recognition capability of radar-based systems and the susceptibility of visual-based systems to adverse weather conditions\cite{8846214},  
the aforementioned single-modality low-altitude target recognition systems \cite{11106829, 8239598, 7508914, 11083683, 11213400, 9349620} possess  inherent limitations.  
To address these shortcomings, recent research has studied the radar-vision fusion system to realize more reliable target recognition.
For example, 
V.~Mehta~\emph{et~al.} utilize  FMCW radar and  Pan-Tilt-Zoom (PTZ) camera to realize UAV 
 detection and recognition within a range of $1$ km\cite{10311246}. 
V.~Mehta~\emph{et~al.} also present an edge-deployable UAV recognition system,
which utilizes YOLOv8 for electro-optical/infrared (EO/IR) visual detection and applies an interacting multiple model (IMM) filter to extract radar kinematic features\cite{11257244}. 
H.~Tang~\emph{et~al.} 
project radar traces onto camera frames and combine radar distance and azimuth information with visual images to realize target detection and recognition\cite{Tang_2024}.
These works provide preliminary technical support for multi-modal low-altitude   target recognition.
However, the large-scale deployment of dedicated radar to realize radar-vision fusion low-altitude monitoring in urban environments faces severe constraints.
For instance, the widespread installation of dedicated radar equipment incurs prohibitive hardware and deployment costs \cite{9296833}. Moreover, it is practically infeasible to find suitable radar installation sites for ubiquitous coverage in dense cityscapes.

Fortunately,  the sixth generation (6G) mobile information network will construct
the  integrated sensing and communications (ISAC) system, 
which enables future base stations (BSs) to sense and monitor various environmental information while simultaneously fulfilling their wireless communications functions\cite{9737357, 9040264, 11250835, 9330512,9681870, 9144301}. 
Currently, both the industry and academia are trying to establish the low-altitude target  monitoring network  centered on ISAC system.
For example, J.~Tang~\emph{et~al.}   propose a low-altitude UAV localization and velocity estimation scheme based on spatial smooth tensor decomposition within cooperative ISAC network
\cite{10906066}.
Y.~Huang~\emph{et~al.} propose a low-altitude target localization scheme based on compressed sensing and sparse imaging within  ISAC network\cite{11151696}. 
J.~Wei~\emph{et~al.} propose a refined micro-Doppler feature extraction method for rotor UAV target\cite{11077832}.
J.~Xue~\emph{et~al.}   propose a UAV and bird recognition scheme based on short-time Fourier transform (STFT) and DC-Former network within ISAC system\cite{11069481}. 
However, to the best of the authors' knowledge,
 no existing studies have investigated
 the fusion of ISAC and vision
 for  low-altitude target recognition.

 In this paper, we propose an ISAC and vision fusion framework for fine-grained  low-altitude target recognition. 
The contributions of this paper are summarized as follows.
\begin{itemize}

\item \textcolor{black}{We  utilize  ISAC system to estimate the position of the  low-altitude target. 
	Then we actively adjust the working parameters of the PTZ camera based on the estimated  target position, such that the camera can capture the image of tiny low-altitude target from several hundred meters away.}

\item We employ  STFT to obtain the mD spectrum of the target from the wireless echo signal, and design a conditional generative adversarial network (cGAN)-based denoising network to optimize the quality of the mD spectrum.
Meanwhile, we employ YOLOv11 to detect the low-altitude target from visual image, and then  crop the small-sized feature image of the target from the original image.

\item We  design a fine-grained low-altitude target recognition network with MobileViT, which can fuse the optimized mD spectrum and the cropped feature image to distinguish the subcategory of low-altitude target.

\item We generate a joint ISAC and vision dataset (JIVD) for low-altitude target monitoring based on AirSim and Wireless InSite, which includes diverse target subcategories, scenarios, and weather conditions.
The effectiveness and superiority of the  proposed    scheme   have been demonstrated by simulation results.

\end{itemize}

The remainder of this paper is organized as follows.
In Section \RNum{2},
we propose the
ISAC and vision fusion enabled 
fine-grained low-altitude target recognition framework.
In Section \RNum{3},
we provide the workflow for obtaining ISAC system echo signals, estimating low-altitude target parameters, adjusting PTZ camera working  parameters, and acquiring low-altitude target images.
In Section~\RNum{4},
 we propose a low-altitude target mD spectrum extraction and denoising enhancement algorithm.
In Section~\RNum{5}, we utilize YOLOv11 to preprocess the visual image, 
and then design a fine-grained low-altitude target recognition network with MobileViT. 
Simulation results and conclusions are given in Section~\RNum{6} and Section~\RNum{7}.

\emph{Notation}: Lower-case and upper-case boldface letters $\mathbf{a}$ and $\mathbf{A}$ denote a vector and a matrix; $\mathbf{a}^*$, $\mathbf{a}^T$ and $\mathbf{a}^H$ denote the conjugate, transpose and conjugate transpose of $\mathbf{a}$, respectively; $\mathbf{A}(i,j)$ denotes the $(i,j)$-th element of the matrix $\mathbf{A}$; $\mathbb{C}^{M \times N}$ and $\mathbb{R}^{M \times N}$ represent the spaces of $M \times N$ complex-valued and real-valued matrices, respectively; $\otimes$, $\odot$ and $\circ$ denote the Kronecker product, Hadamard product and function composition, respectively; $\lfloor \cdot \rfloor$ denotes the floor operation; $\mathbb{E}[\cdot]$ represents the statistical expectation; $\mathcal{F}(\cdot)$ and $\mathcal{F}^{-1}(\cdot)$ denote the fast Fourier transform (FFT) and inverse fast Fourier transform (IFFT) operations.

\section{The Proposed ISAC and Vision Fusion Framework for Low-Altitude Target Recognition}

\begin{figure}[!t]
	\centering
	\includegraphics[width=76mm]{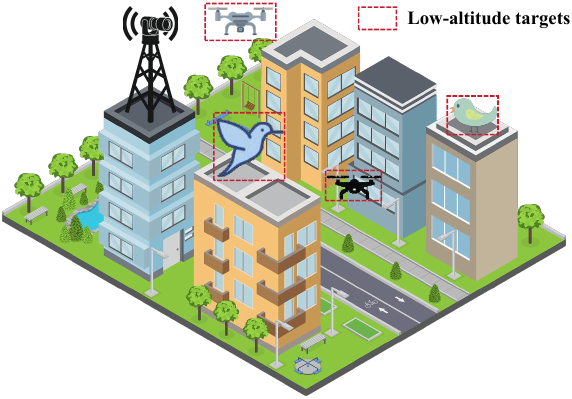}
	\caption{Fine-grained low-altitude target recognition scenario based on ISAC and vision fusion.}
	\label{scene}
\end{figure}

In this section, 
we   propose the
ISAC and vision fusion enabled 
fine-grained low-altitude target recognition framework.

\begin{figure*}[!t]
	\centering
	\includegraphics[width=180mm]{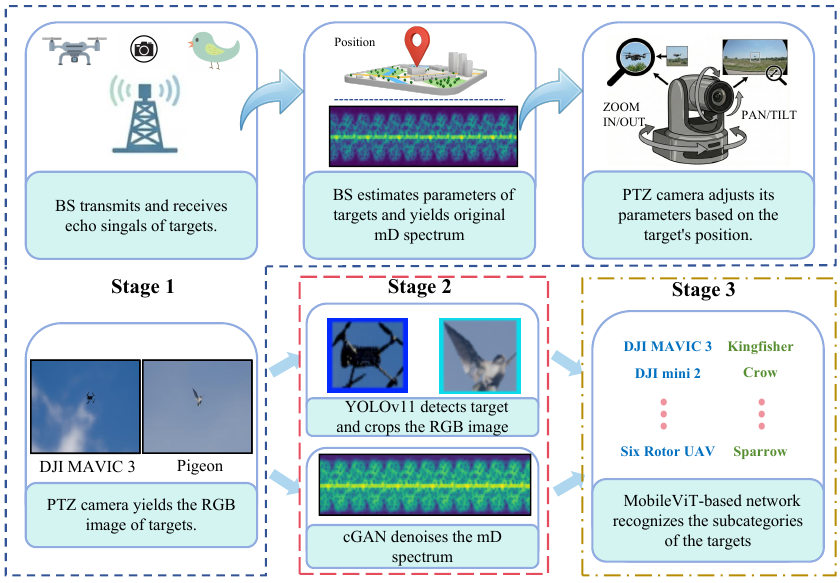}
	\caption{The proposed ISAC and vision fusion framework for fine-grained low-altitude target recognition.}
	\label{Framework2}
\end{figure*}

\subsection{ISAC and Vision Fusion System Model}

The considered low-altitude monitoring  system
 includes one ISAC BS and one PTZ camera, as depicted in Fig.~\ref{scene}.
Generally, the ISAC BS transmits sensing signals and then processes the backscattered echo signals to estimate the position  of the low-altitude target. 
 Simultaneously, the PTZ camera dynamically adjusts its orientation and focal length guided by the estimated target position,
 and then the PTZ camera can capture the image of tiny low-altitude target from several hundred meters away. 
The low-altitude targets of interest in this work primarily comprise UAVs and birds. 
Since the subcategories of UAVs and birds exhibit diversity, we design the ISAC and vision fusion system to recognize fine-grained UAVs and birds by fusing wireless features and visual features.

The ISAC BS operates in  mmWave frequency bands with orthogonal frequency division multiplexing (OFDM) modulation to realize both wireless communications and low-altitude target monitoring. Suppose that  ISAC system emits   OFDM signals with $M$ subcarriers,
where the  lowest  frequency and the subcarrier  interval  are $f_0$ and $\Delta f$, respectively. 
Then the transmission bandwidth is $W=M\Delta f$, and 
the frequency of the $m$-th subcarrier  is $f_m=f_0+m\Delta f$,  $m=0,1,...,M-1$.
We consider that one OFDM
frame contains $N$ consecutive OFDM symbols. 
The full  $N$ symbols are utilized  for target recognition, while the first $N_{0}$ symbols are utilized for target position and velocity estimation.
The time interval between the adjacent  symbols is  $T_s$\cite{4570206}. 

The ISAC BS employs massive 
multiple input multiple output (MIMO)
 arrays with one {hybrid unit uniform rectangular array (HU-URA)} and one {radar unit uniform rectangular array (RU-URA)}. 
 The HU-URA and RU-URA
 are responsible for transmitting sensing detection signals and receiving target echo signals, respectively.
Both the arrays are vertically mounted on the 2D plane $y=0$. The HU-URA and RU-URA contain $N_H = N_{H}^x \times N_{H}^z$ and $N_R = N_{R}^x \times N_{R}^z$ antenna elements, respectively. The inter-element spacing along the $x$-axis and $z$-axis is  $d = \lambda/2$, where $\lambda$ denotes the carrier wavelength.

Let us assume that   BS is located at the origin of the coordinate system,
and  PTZ camera is co-located with  BS.
Moreover, the PTZ camera possesses three controllable degrees of freedom, i.e., the pan angle $\psi$, the tilt angle $\vartheta$, and the focal length $f$. 
Among these working parameters, 
 pan angle and tilt angle determine the optical axis direction of the camera,
while  focal length controls the camera's field of view (FOV). 
Let $\omega_{R}$ and $\omega_{L}$ denote the right and left azimuth angular boundaries of the camera at $f$. Then the  FOV can be formulated as
\begin{equation}
	FOV = \omega_{R} - \omega_{L}.
\end{equation}
Let us adopt the spherical coordinate system $(r, \theta, \phi)$ to describe the 3D spatial relationship. Here, $r \ge 0$ denotes the radial distance, while $\theta \in [30^\circ, 150^\circ]$ and $\phi \in [-60^\circ, 60^\circ]$ represent the azimuth angle and elevation angle, respectively. With the  BS and the PTZ camera co-located at the origin, the effective monitoring region $\mathcal{S}$ can be defined as
\textcolor{black}{\begin{equation}
\begin{split}
\mathcal{S} = \big\{ (r, \theta, \phi) \mid {} & r_{\min} \le r \le r_{\max}, \theta_{\min} \le \theta \le \theta_{\max}, \\
& \kern 73pt \phi_{\min} \le \phi \le \phi_{\max} \big\}.
\end{split}
\end{equation}}We assume that all the low-altitude targets are located within this monitoring area.

\subsection{The Proposed ISAC and Vision Fusion Framework for Fine-Grained Low-Altitude Target Recognition}

As illustrated in Fig.~\ref{Framework2}, 
the proposed ISAC and vision fusion framework for fine-grained low-altitude target recognition
includes three  stages, which are introduced as follows.

\textbf{1) BS Sensing and Image Acquisition.}
In the first stage,
the HU-URA of  BS covers the entire sensing space through wide beamforming and emits the sensing detection signals,  which will be scattered by low-altitude target and cause echo signals.
Then the RU-URA of BS will receive these target echo signals. 
We utilize 2D-FFT and minimum variance distortionless response (MVDR) algorithms to estimate the position and velocity parameters of the  low-altitude target from the echo signals as $(\hat{r}, \hat{\theta}, \hat{\phi}, \hat{v}_{r})$.
Meanwhile, we also implement STFT on the echo  signals to obtain the original mD spectrum of the low-altitude target. 
Subsequently, we actively adjust the working parameters $\{ \psi, \vartheta, f\}$ of the PTZ camera based on the estimated $(\hat{r}, \hat{\theta}, \hat{\phi}, \hat{v}_{r})$.
Then the camera can capture the proper RGB image of tiny low-altitude target from several hundred meters away.

\textbf{2) Feature Extraction and Enhancement.}
In the second stage, we design a cGAN-based denoising network  to optimize the quality of  the original mD spectrum obtained by ISAC BS \cite{goodfellow2014generative}. 
Simultaneously, we 
employ YOLOv11 network to detect the low-altitude target from the captured visual image \cite{khanam2024yolov11}.
Based on the detection results,
we  crop the small-sized feature image of the target from the original image.
The small-sized feature image effectively removes  environmental background interference and makes the visual modality more focused on  target recognition.

\textbf{3) Dual-Modal Feature Fusion and Target Recognition.}
In the third stage, we leverage MobileViT \cite{mehta2021mobilevit} and other  deep learning techniques to facilitate interaction and fusion of the preprocessed wireless feature and visual feature. 
By integrating the micro-motion information embedded in the optimized mD spectrum with the rich texture, color, and shape attributes from the RGB visual image, we extract the critical features essential for fine-grained target recognition. Through iterative training on the generated dataset, we can realize   high-precision fine-grained low-altitude target recognition.


%
%
%
%
%
%

\section{\textcolor{black}{
		Low-Altitude Target Parameter Estimation and PTZ Camera Image Acquisition}}

In this section, we utilize  ISAC BS to estimate the position and velocity information of the  low-altitude target.
Then we utilize the estimated  target position to control the PTZ camera for precise target image acquisition.

\subsection{Echo Signals of Low-Altitude Target}

Let us consider a low-altitude volumetric target with $K$ scattering points. 
For the $k$-th scattering point, let $(r_{n,k}, \theta_{n,k}, \phi_{n,k})$ denote its position at the $n$-th OFDM symbol, $n=0,1,...,N-1$, and $k=0,...,K-1$. Then the  sensing echo channel of the  $k$-th scattering  point on the $m$-th subcarrier of the $n$-th OFDM symbol can be represented as \cite{11159257}
\begin{equation}
	\label{deqn_ex1a}
	\begin{split}
		\mathbf{H}_{n,m,k} ={}& \alpha_{n,k} 
		e^{-j \frac{4\pi f_m r_{n,k}}{c}}
\mathbf{a}_{R}(\Psi_{n,k},\Omega_{n,k})
		\mathbf{a}^T_{H}(\Psi_{n,k},\Omega_{n,k}),
	\end{split}
\end{equation}
where $\alpha_{n,k}$ denotes the complex channel fading factor, $\Psi_{n,k} = \cos \phi_{n,k} \cos \theta_{n,k}$ and $\Omega_{n,k} = \sin \phi_{n,k}$ represent the spatial directions. 
Moreover, $\mathbf{a}_{R} \in \mathbb{C}^{N_R \times 1}$ and $\mathbf{a}_{H} \in \mathbb{C}^{N_H \times 1}$ represent the array steering vectors for the RU-URA and HU-URA, respectively \cite{7523373}. 
Based on the URA structure, the array vectors are decomposed by the Kronecker product $\otimes$ as
\begin{align}
	\mathbf{a}_{R}(\Psi_{n,k},\Omega_{n,k}) &= \mathbf{a}_{R}^x(\Psi_{n,k}) \otimes \mathbf{a}_{R}^z(\Omega_{n,k}), \\
	\mathbf{a}_{H}(\Psi_{n,k},\Omega_{n,k}) &= \mathbf{a}_{H}^x(\Psi_{n,k}) \otimes \mathbf{a}_{H}^z(\Omega_{n,k}).
\end{align}
The steering vectors along the x-axis and z-axis follow the standard uniform linear array (ULA) format, i.e.,
\begin{align}
	\!\!\mathbf{a}_{R}^x(\Psi)&\!=\![1,e^{j\frac{2\pi f_0d\Psi}{c}},...,e^{j\frac{2\pi f_0d\Psi}{c}(N_{R}^x-1)}]^T \!\!\in\! \mathbb{C}^{N_{R}^x\times 1},\\
	\!\!\mathbf{a}_{R}^z(\Omega)&\!=\! [1,e^{j\frac{2\pi f_0d\Omega}{c}},...,e^{j\frac{2\pi f_0d\Omega}{c}(N_{R}^z-1)}]^T \!\! \in\! \mathbb{C}^{N_{R}^z\times 1},\\
	\!\!\mathbf{a}_{H}^x(\Psi)&\!=\![1,e^{j\frac{2\pi f_0d\Psi}{c}},...,e^{j\frac{2\pi f_0d\Psi}{c}(N_{H}^x-1)}]^T \!\!\in\! \mathbb{C}^{N_{H}^x\times 1},\\
	\!\!\mathbf{a}_{H}^z(\Omega)&\!=\! [1,e^{j\frac{2\pi f_0d\Omega}{c}},...,e^{j\frac{2\pi f_0d\Omega}{c}(N_{H}^z-1)}]^T \!\! \in\! \mathbb{C}^{N_{H}^z\times 1}.
\end{align}
Based on (3), the  sensing echo channel of the  low-altitude target with $K$ scattering points on the $m$-th subcarrier of the $n$-th OFDM symbol can be represented as \cite{9947033}
\begin{equation}
	\label{deqn_ex1a}
	\begin{split}
		\mathbf{H}_{n,m} 
		&= \sum_{k=0}^{K-1} \mathbf{H}_{n,m,k}. \\
	\end{split}
\end{equation}

We adopt the generalized step-chirp (GSC) sequence design \cite{10634315} for the transmitting beamforming of HU-URA to realize efficient low-altitude wide-area coverage. 
The wide-area beamforming vector $\mathbf{W}_H \in \mathbb{C}^{N_H \times 1}$ can be  denoted as
\begin{align}
	\mathbf{W}_H = 
	\mathbf{W}_{H}^x \otimes \mathbf{W}_{H}^z.
\end{align}
Here $\mathbf{W}_{H}^x$ and $\mathbf{W}_{H}^z$ control the azimuth and elevation beamwidths,  whose  specific expressions can be found in \cite{10634315}.

Based on (11), the transmission signal on the $m$-th subcarrier of the
$n$-th OFDM symbol is 
\begin{equation}
	\begin{split}
		\begin{aligned}
			\label{deqn_ex1a}
			{\mathbf{x}}_{n,m} =  \sqrt{{P_t}/{N_H}}\mathbf{W}_Hs_{n,m},
		\end{aligned}
	\end{split}
\end{equation}
where $P_t$ is the transmission power of BS,  and $s_{n,m}$ is the sensing detection signal. 

Then the sensing echo signals on the $m$-th subcarrier of the
$n$-th  symbol received by RU-URA can be represented
as
\begin{equation}
	\begin{split}
		\begin{aligned}
			\label{deqn_ex1a}
			\mathbf{y}_{n,m} =  \mathbf{H}_{n,m} {\mathbf{x}}_{n,m}^* + \mathbf{n}_{n,m},
		\end{aligned}
	\end{split}
\end{equation}
where $\mathbf{n}_{n,m}$ is  zero-mean additive   Gaussian   noise  with variance  $\sigma^2$.
Then we  erase the influence of $s_{n,m}$ and obtain
\begin{equation}
	\begin{split}
		\begin{aligned}
			\label{deqn_ex1a}
			\mathbf{\check{y}}_{n,m} = {\mathbf{y}}_{n,m}/s_{n,m}.
		\end{aligned}
	\end{split}
\end{equation}Note that $\mathbf{\check{y}}_{n,m} \in \mathbb{C}^{N_R \times 1}$ denotes the received signal vector across all antenna elements. Hence, $\mathbf{\check{y}}_{n,m}$  can be  reshaped into a 2D spatial matrix $\mathbf{Y}_{n,m} \in \mathbb{C}^{N_{R}^x \times N_{R}^z}$, expressed as
\begin{equation}
	\mathbf{Y}_{n,m} \in \mathbb{C}^{N_{R}^x \times N_{R}^z}, \quad \text{s.t.} \quad \mathrm{vec}(\mathbf{Y}_{n,m}) = \mathbf{\check{y}}_{n,m}.
\end{equation}Consequently, the total received echo  signals of the RU-URA across all OFDM symbols and all subcarriers can be represented as a 4D tensor, i.e.,
\begin{equation}
	\mathcal{Y} = \{ \mathbf{Y}_{n,m} \}_{n=0, m=0}^{N-1, M-1}, \quad \mathcal{Y} \in \mathbb{C}^{N \times M \times N_{R}^x \times N_{R}^z}.
\end{equation}

\subsection{Target Position and Velocity Estimation}
Note that the velocity of the target is implicitly embedded within the sequence of position $(r_{n,k}, \theta_{n,k}, \phi_{n,k})$ for $n=0,1,...,N-1$. Hence, $\mathcal{Y}$ contains not only the positional information of the target but also its velocity information. 
Here, we utilize the ISAC BS to estimate the position and radial velocity of the low-altitude target's centroid.
We first obtain a coarse range and velocity estimation by 2D-FFT. 
Let $\mathbf{Y}_{RD} \in \mathbb{C}^{N_0 \times M}$ denote the echo signal extracted from the $(0,0)$-th antenna element in $\mathcal{Y}$. The range-Doppler spectrum $\mathbf{G}_{RD}$ is obtained via 2D-FFT as
\begin{equation}
	\mathbf{G}_{RD} = \mathcal{F}((\mathcal{F}^{-1} ( \mathbf{Y}_{RD}, M, 2 )), L_v, 1),
\end{equation}
where $\mathcal{F}^{-1} ( \cdot, M, 2 )$ represents the $M$-point IFFT operation along the second dimension, and $\mathcal{F}\left( \cdot, L_v, 1 \right)$ represents the $L_v$-point FFT operation along the first dimension.
The coarse estimated range $\hat{r}_c$ and velocity $\hat{v}_c$ are obtained  by the peak indices $(n_{d}, n_{r})$ of the amplitude spectrum $|\mathbf{G}_{RD}|$, i.e.,
\begin{equation}
	\hat{r}_c = n_{r} \Delta r, \quad \hat{v}_c = n_{d} \Delta v,
\end{equation}
where $\Delta r = c/2W$ and $\Delta v = \lambda / (2 L_v T_s)$ represent the range and velocity resolutions, respectively.

Then we employ MVDR for range refinement. Let $\mathbf{Y}_{r} \in \mathbb{C}^{M \times N_{R}}$ denote the echo matrix extracted at the $0$-th OFDM symbol. The covariance matrix can be computed as
\begin{equation}\hat{\mathbf{R}}_r = \frac{1}{N_{R}} \mathbf{{Y}}_{r}\mathbf{{Y}}_{r}^H.\end{equation}
Then the range MVDR spectrum can be  formulated as
\begin{equation}
	P_{r}(r) = \frac{1}{\mathbf{a}^H(r) \hat{\mathbf{R}}_r^{-1} \mathbf{a}(r)},
\end{equation}
where 
$\mathbf{a}(r) = [1, e^{-j\frac{4\pi \Delta f r}{c}}, \dots, e^{-j\frac{4\pi \Delta f r}{c}(M-1)}]^T$ is range steering vector.
The precise range $\hat{r}$ \textcolor{black}{can  be} estimated by searching the peak of $P_{r}$ within a local window $[\hat{r}_{c} - \delta_r, \hat{r}_{c} + \delta_r]$, where  $\delta_r$ represents the range window width.

Similarly, let $\mathbf{Y}_{v} \in \mathbb{C}^{N_0 \times N_{R}}$ denote the echo matrix extracted at the $0$-th subcarrier. The  covariance matrix and velocity MVDR spectrum can be computed as
\begin{align}
	\hat{\mathbf{R}}_v &= \frac{1}{N_{R}} \mathbf{{Y}}_{v}\mathbf{{Y}}_{v}^H, \\
	P_{v}(v) &= \frac{1}{\mathbf{a}^H(v) \hat{\mathbf{R}}_v^{-1} \mathbf{a}(v)},
\end{align}
where 
$\mathbf{a}(v) = [1, e^{j\frac{4\pi v T_s}{\lambda}}, \dots, e^{j\frac{4\pi v T_s}{\lambda}(N_0-1)}]^T$ is velocity steering vector.
The precise velocity $\hat{v}$ is estimated by searching for the peak of $P_{v}$ within a local window $[\hat{v}_{c} - \delta_v, \hat{v}_{c} + \delta_v]$, where  $\delta_v$ represents the velocity window width.

Next, we  \textcolor{black}{need to obtain} a coarse $ \hat{\theta}_c $ by FFT. Let $\mathbf{Y}_{a} \in \mathbb{C}^{N_{R}^x \times N_{R}^z}$ denote the spatial echo matrix extracted on the $0$-th subcarrier at the $0$-th OFDM symbol. The azimuth profile can be obtained by
\begin{equation}
	\mathbf{g_\theta} = 
	\frac{1}{N_{R}^z} \sum_{n_{R}^z=0}^{N_{R}^z-1} \left| \mathcal{F}\left( \mathbf{Y}_{a}(:,n_{R}^z), L_a, 1 \right) \right|.
\end{equation}
Then  $\hat{\theta}_{c}$ is determined by the peak index of $\mathbf{g}_{\theta}$, which can be expressed as
$\hat{\theta}_{c} = \arcsin \left( \frac{\arg \max \mathbf{g}_{\theta} - L_a/2}{L_a/2} \right) \frac{180}{\pi}.$

Let us extract $\mathbf{Y}_{\theta} \in \mathbb{C}^{N_{R}^x \times N_{R}^zM}$ at the $0$-th OFDM symbol. The azimuth covariance matrix and  MVDR spectrum are computed as
\begin{equation}\hat{\mathbf{R}}_{\theta} = \frac{1}{N_{R}^zM} \mathbf{Y}_{\theta}\mathbf{Y}_{\theta}^H,
\end{equation}
\begin{equation}P_{\theta}(\theta) = \frac{1}{\mathbf{a}^H(\theta) \hat{\mathbf{R}}_{\theta}^{-1} \mathbf{a}(\theta)},
\end{equation}
where the azimuth steering vector is defined as $\mathbf{a}(\theta) = [1, e^{j \pi \sin \theta}, \dots, e^{j \pi (N_{R}^x-1) \sin \theta}]^T.$ The precise azimuth $\hat{\theta}$ is estimated by searching for the peak of $P_{\theta}$ within a local window $[\hat{\theta}_{c} - \delta_{\theta}, \hat{\theta}_{c} + \delta_{\theta}]$, where $\delta_{\theta}$ represents the search scope.

\textcolor{black}{Since the RU-URA structure exhibits symmetry between the horizontal and vertical dimensions, the estimation of the elevation angle $\phi$ could follow the same procedure as that of the azimuth. We obtain 
	the precise elevation   estimated $\hat{\phi}$ by performing the coarse-FFT and fine-MVDR  along the $N_{R}^z$ dimension.} 
Eventually, we have estimated the position and velocity parameters of the low-altitude target  as $(\hat{r}, \hat{\theta}, \hat{\phi}, \hat{v}_{r})$.

\subsection{PTZ Camera Active Control and Image Acquisition}

Based on the estimated target parameters $(\hat{r}, \hat{\theta}, \hat{\phi}, \hat{v}_{r})$ from ISAC BS, we  can actively control the PTZ camera to
obtain the effective RGB visual image of the tiny low-altitude target from  hundreds of meters away.


Let $\boldsymbol{\omega} = [\psi, \vartheta, \gamma]^T$ denote the desired Euler angles,  which includes yaw, pitch, and roll of the camera. 
Based on  estimated  $\hat{\phi}$ and $\hat{\theta}$, the control law is defined as \begin{equation}\psi = \hat{\phi}, \quad \vartheta = \hat{\theta}, \quad \gamma = 0,\end{equation}where the roll angle $\gamma$ is fixed at zero to maintain a stable horizon. The Euler angles are then converted into a unit quaternion $\mathbf{q} = [q_w, q_x, q_y, q_z]^T$ as
\begin{equation}
\mathbf{q} = 
\begin{bmatrix}
q_w \\
q_x \\
q_y \\
q_z
\end{bmatrix}
=
\begin{bmatrix}
\cos\left(\frac{\psi}{2}\right)\cos\left(\frac{\vartheta}{2}\right) \\
\cos\left(\frac{\psi}{2}\right)\sin\left(\frac{\vartheta}{2}\right) \\
\sin\left(\frac{\psi}{2}\right)\cos\left(\frac{\vartheta}{2}\right) \\
-\sin\left(\frac{\psi}{2}\right)\sin\left(\frac{\vartheta}{2}\right)
\end{bmatrix}.
\end{equation}
The generated quaternion $\mathbf{q}$ serves as the direct control input for the PTZ camera to ensure precise angular alignment with the low-altitude target.

Then we adjust the FOV to realize adaptive zooming for low-altitude targets. Let $L_{obj}$ denote the physical width of the target in the real world and $L_{cls}$ denote the required pixel width for the effective recognition. Given the total image width $W_{img}$, the required physical coverage width $W_{cls}$  can be derived as
\begin{equation}W_{cls} = \frac{W_{img}}{L_{cls}} \cdot L_{obj}.
\end{equation}

Based on the pinhole imaging model, the relationship between the azimuth FOV and the coverage width $W_{cls}$ at distance $\hat{r}$ is given by geometrical similarity as
\begin{equation}\tan\left(\frac{FOV}{2}\right) = \frac{W_{cls}}{2\hat{r}}.\end{equation}
Substituting (28) into (29), the adaptive FOV control law can be formulated as
\begin{equation}FOV = 2 \arctan \left( \frac{W_{img} \cdot L_{obj}}{2 \hat{r} \cdot L_{cls}} \right).\end{equation}

\textcolor{black}{It should be noted that the derived ${\mathbf{q}}$ and $FOV$ is mathematically equivalent to PTZ camera's working parameters $\{\psi, \vartheta, f\}$, which implies that to adjiust ${\mathbf{q}}$ and $FOV$ is fundamentally identical to adjusting  $\{\psi, \vartheta, f\}$.
Hence, based on $\mathbf{q}$ and $FOV$, we can align the main optical axis of the PTZ camera with the spatial direction of the low-altitude target, and the PTZ camera's perspective matches the distance and geometric dimensions of the target. Then we can capture the effective RGB  image of the  tiny low-altitude target from hundreds of meters away.}

\section{\textcolor{black}{Micro-Doppler Spectrum Extraction and Denoising Enhancement from ISAC BS}}
In this section, we obtain the mD spectrum observed by the ISAC BS and perform denoising enhancement by cGAN.

\subsection{Micro-Doppler Spectrum Extraction}
Based on the estimated angles $\hat{\theta}$ and $\hat{\phi}$, we construct the received beamforming weight matrix $\hat{\mathbf{W}}_{R} \in \mathbb{C}^{N_{R}\times 1} $ as
\begin{equation}
\hat{\mathbf{W}}_{R} = \frac{1}{\sqrt{N_R}} \mathbf{a}_{R}(\hat{\theta}, \hat{\phi}).
\end{equation}
The  signal after received beamforming can be  obtained as ${y}_{n,m}^{BF} = \hat{\mathbf{W}}_{R}^{H} \mathbf{\check{y}}_{n,m} $. Then the  echo signal matrix after received beamforming can be expressed as
\begin{equation}
\!\!\mathbf{Y}_{\!BF} = \!\!
\begin{bmatrix}
y_{0,0}^{BF} & y_{0,1}^{BF} & \cdots & y_{0, M-1}^{BF} \\
y_{1,0}^{BF} & y_{1,1}^{BF} & \cdots & y_{1, M-1}^{BF} \\
\vdots & \vdots & \ddots & \vdots \\
y_{N-1,0}^{BF} & y_{N-1,1}^{BF} & \cdots & y_{N-1, M-1}^{BF}
\end{bmatrix} \!\! \in \! \mathbb{C}^{N \times M}.
\end{equation}

Next, we apply STFT on $\mathbf{Y}_{BF}$ to capture the time-varying micro-motion features of low-altitude targets with a sliding window of length $L_{win}$ and a step size $A_{step}$. 
Let $\mathbf{Y}_{BF}^{(q)}$ denote the $q$-th signal segment along the symbol dimension, $q = 0,1,...,Q-1$. The total number of segments $Q$ is
\begin{equation}
Q = \left\lfloor \frac{N - L_{win}}{A_{step}} \right\rfloor + 1.
\end{equation}
Then the $\mathbf{Y}_{BF}^{(q)}$ can be extracted as
\begin{equation}
\mathbf{Y}_{BF}^{(q)} =  \mathbf{Y}_{BF}(q \cdot A_{step}:q \cdot A_{step} + L_{win} - 1, :).\end{equation}
The range-Doppler map for the $q$-th segment is computed as
\begin{equation}
\!\!\!\mathbf{G}^{(q)}_{mD} \! =\! \mathcal{F} \left( \left( \mathcal{F}^{-1} \left( \mathbf{Y}_{BF}^{(q)} \odot \mathbf{W}_{hann}, M, 2\right ) \right), L_{win}, 1 \right),
\end{equation}
where  $\mathbf{W}_{hann}$ is the Hanning window function.
Subsequently, we obtain the $q$-th column of the raw mD spectrum $\mathbf{S}_{raw} \in \mathbb{R}^{L_{win} \times Q}$ by summing the magnitude of $\mathbf{G}^{(q)}_{mD}$ over the range dimension as 
\begin{equation}
\mathbf{S}_{raw}(l,q) \!=\! \sum_{m=0}^{M-1} \left| \mathbf{G}^{(q)}_{mD}(l,m) \right|, \quad l = 0, \dots, L_{win}-1.
\end{equation}

Define $v_{min} = \min(\mathbf{S}_{raw})$ as the minimum value of $\mathbf{S}_{raw}$. Then we could construct the padded spectrum $\mathbf{S}_{pad} \in \mathbb{R}^{3L_{win} \times Q}$ as
\begin{equation}
\mathbf{S}_{\text{pad}}(l,q) =
\begin{cases}
\mathbf{S}_{\text{raw}}(\,l - L_{\text{win}},\, q), & L_{\text{win}} \le l < 2L_{\text{win}}, \\
v_{\min}, & \text{otherwise}.
\end{cases}
\end{equation}
The energy center index $l_c$ is obtained by maximizing the accumulated energy along the time dimension, i.e.,
\begin{equation}
l_c = \arg \max_{l} \sum_{q=0}^{Q-1} \mathbf{S}_{pad}(l,q).
\end{equation}
Next, we crop a fixed window centered at $l_c$ to extract the aligned mD spectrum $\mathbf{S}_{aln} \in \mathbb{R}^{L_{win} \times Q}$, as
\begin{equation}
\mathbf{S}_{aln}(l,q) = \mathbf{S}_{pad}(l_c - \lfloor L_{win}/2 \rfloor + l , q).
\end{equation}

\begin{figure}[!t] 
    \centering

   \subfloat[]{
        \includegraphics[width=0.5\linewidth]
        {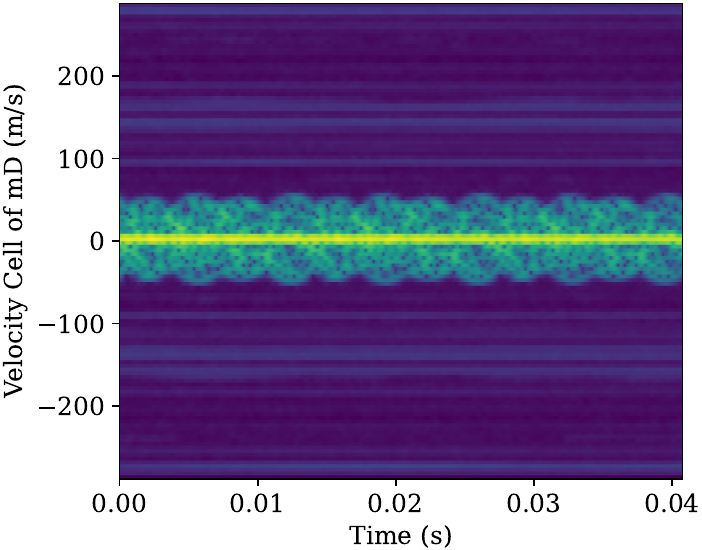}
        \label{fig:sub_a}
    }
    \subfloat[]{
        \includegraphics[width=0.5\linewidth]
        {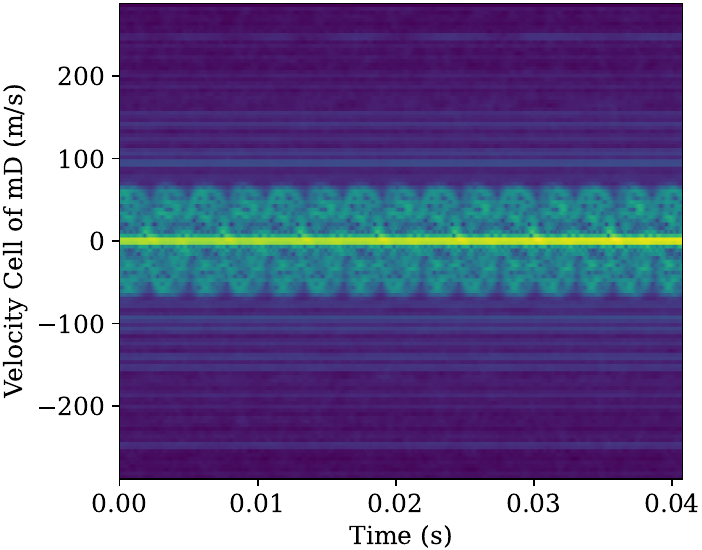}
        \label{fig:sub_b}
    }

  \subfloat[]{
        \includegraphics[width=0.5\linewidth]
        {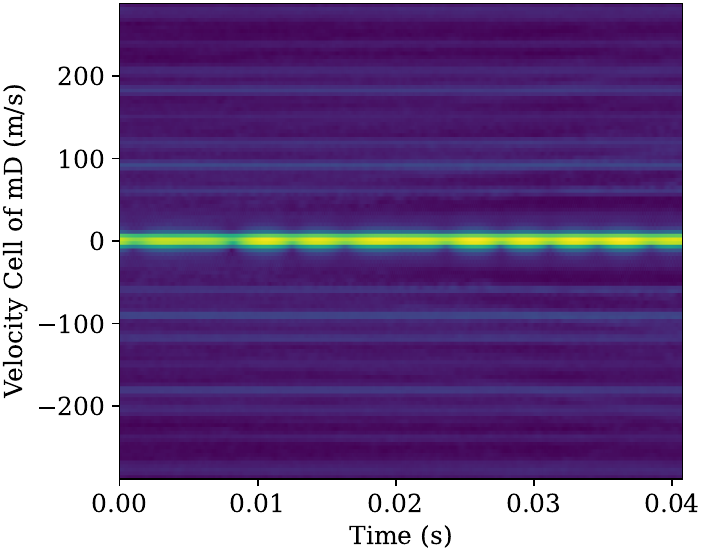}
        \label{fig:sub_c}
    }
   \subfloat[]{
        \includegraphics[width=0.5\linewidth]
        {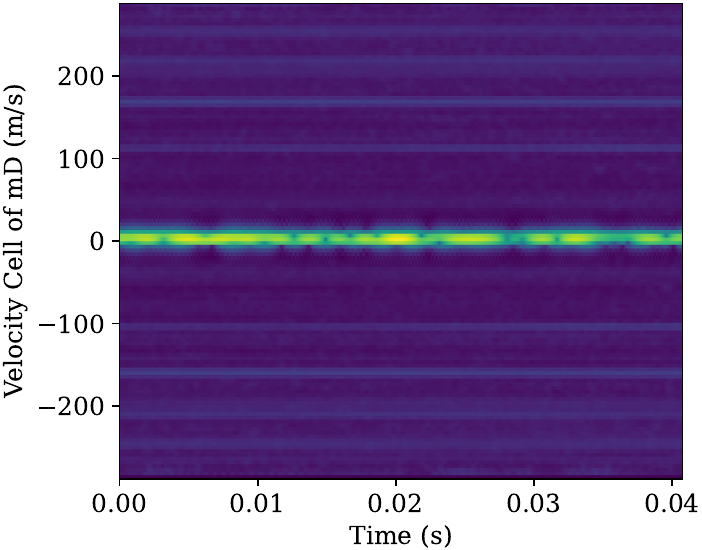}
        \label{fig:sub_d}
    }

    \caption{mD spectrum of four different targets. (a) DJI MAVIC 3. (b) DJI MINI 2. (c) Kingfisher. (d) Crow.}
    \label{mD}
\end{figure}

For visualization and subsequent denoising, we convert $\mathbf{S}_{aln}$ into an RGB image $\mathbf{X}_{mD} \in \mathbb{R}^{H_r \times W_r \times 3}$.
Fig.~\ref{mD} illustrates that the obtained mD spectrum exhibit distinct micro-motion signatures for different subcategories of low-altitude targets. 
Specifically, Fig.~\ref{mD}(a) and Fig.~\ref{mD}(b) display consistent and periodic micro-motion patterns of DJI MAVIC 3 and DJI MINI 2, which are attributed to the high-speed rotation of the rotor blades of UAVs. Furthermore, discernable differences are also evident among 
different subcategories UAV models, which are attributed to distinct rotor rotation frequencies and structural configurations. 
In contrast, the mD spectrum of the birds in Fig.~\ref{mD}(c) and Fig.~\ref{mD}(d) demonstrate more irregular and fluctuating patterns. The flapping motion of the Kingfisher and Crow produces lower-frequency modulation components with varying time intervals, which are fundamentally different from the rigid rotation of UAVs. Moreover, difference between Kingfisher and Crow also exists because of different flapping frequencies and body shapes.

\subsection{Micro-Doppler Spectrum Denoising Enhancement}
\begin{figure*}[!t]
	\centering
	\includegraphics[width=180mm]{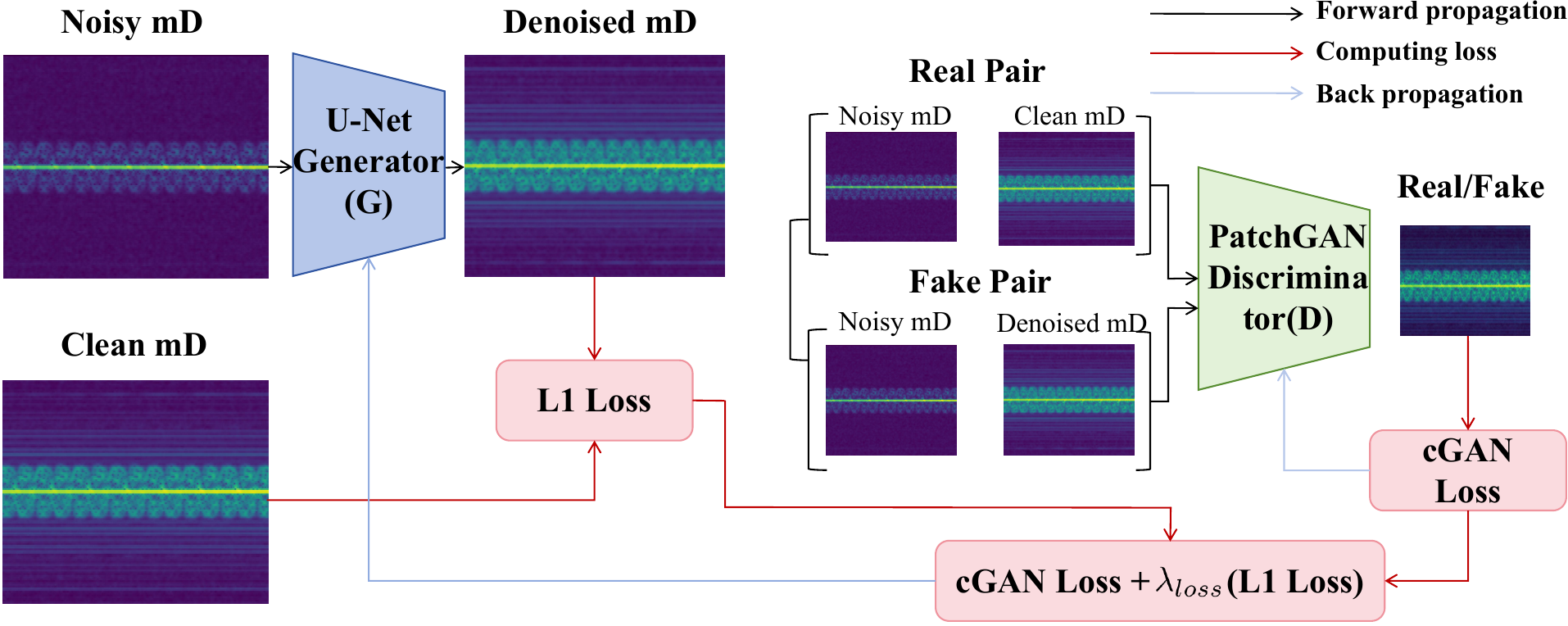}
	\caption{cGAN-based denoising network.}
	\label{cGAN}
\end{figure*}

In practical applications, the received echo signals are inevitably corrupted by noise. Consequently, under low signal-to-noise ratio (SNR) conditions, the mD features become obscured and difficult to distinguish. Let us denote SNR as
\begin{equation} \label{eq:snr_def} 
	\text{SNR} = \frac{\mathbb{E}\left[ \left\| \mathbf{H}_{n,m} \mathbf{x}_{n,m}^* \right\|^2 \right]}{\mathbb{E}\left[ \left\| \mathbf{n}_{n,m} \right\|^2 \right]}. 
\end{equation}
We design a cGAN-based denoising network to denoise the noisy mD spectrum. As illustrated in Fig.~\ref{cGAN}, cGAN consists of a generator $G$ and a discriminator $D$. The generator $G$ aims to map the noisy mD spectrum image $\mathbf{X}_{mD}^{noisy}$ to a clean spectrum image $\mathbf{X}_{mD}^{clean}$.
To preserve low-level features while extracting high-level semantic information, we  adopt a U-Net architecture with symmetric encoder-decoder layers \cite{ronneberger2015u}. 
Define $\mathcal{E}_i$ and $\mathcal{D}_i$ as the $i$-th downsampling block of the encoder and the $i$-th upsampling block of the decoder, $\quad i=1,\dots,I$. 
The encoder maps the input $\mathbf{X}_{mD}^{noisy}$ into a latent feature representation through $I$ hierarchical layers. The output of the $i$-th encoder layer, $\mathbf{e}_i$, is expressed as
\begin{equation}\mathbf{e}_i = \sigma_{ReLU}(\text{BN}(\text{Conv}(\mathbf{e}_{i-1}))),\end{equation}
where $\mathbf{e}_0 = \mathbf{X}_{mD}^{noisy}$, $\text{Conv}$ denotes the convolution operation, $\text{BN}$ is batch normalization, and $\sigma_{ReLU}$ is the ReLU activation function. The decoder fuses the features from the previous decoder layer and the corresponding encoder layer via channel-wise concatenation. 
The output of the $j$-th decoder layer is \begin{equation}\mathbf{d}_j \!=\! \sigma_{ReLU}(\text{BN}(\text{Deconv}([\mathbf{d}_{j-1}, \mathbf{e}_{I-j}]))),\!\!\quad j=1,\dots,I-1,\end{equation}
where $\mathbf{d}_0 = \mathbf{e}_I$ represents the bottleneck feature representation extracted by the encoder, $[\cdot, \cdot]$ denotes the concatenation operation, and $\text{Deconv}$ represents transposed convolution. Then the denoised mD spectrum is generated via a Tanh activation as $\hat{\mathbf{X}}_{mD} = \tanh(\text{Conv}(\mathbf{d}_{I-1}))$.

The discriminator $D$ is designed to distinguish between the real pair $(\mathbf{X}_{mD}^{noisy}, \mathbf{X}_{mD}^{clean})$ and the fake pair $(\mathbf{X}_{mD}^{noisy}, \hat{\mathbf{X}}_{mD})$. To capture local high-frequency details, we  adopt a PatchGAN architecture \cite{8100115} in $D$, which maps the input pair to a validity matrix. Each element in the matrix represents the authenticity probability of a specific local image patch. Mathematically, the discriminator is modeled as a cascade of $L_D$ convolutional blocks. Let $f_u(\cdot)$ denote the mapping function of the $u$-th layer, which typically comprises convolution, normalization, and activation operations. The output validity matrix $\mathbf{M}_{out}$ is thus expressed as
\begin{equation}\mathbf{M}_{out} = (f_{L_D-1} \circ f_{L_D-2} \circ \dots \circ f_0) ([\mathbf{X}_{mD}^{noisy}, \mathbf{X}_{mD}^{cand}]),\end{equation}
where $\mathbf{X}_{mD}^{cand} \in \{\mathbf{X}_{mD}^{noisy}, \hat{\mathbf{X}}_{mD}\}$ represents the candidate image, and $\circ$ denotes function composition.

We design the  training objective as a weighted combination of the adversarial loss and the pixel-wise reconstruction loss. \textcolor{black}{Let $\mathcal{L}$ denote the loss function} and define $\mathbf{M}_{out}^{real} = D(\mathbf{X}_{mD}^{noisy}, \mathbf{X}_{mD}^{clean})$ and $\mathbf{M}_{out}^{fake} = D(\mathbf{X}_{mD}^{noisy}, \hat{\mathbf{X}}_{mD})$. The adversarial loss $\mathcal{L}_{cGAN}$  is then formulated as
\begin{equation}
	\begin{split}
		\mathcal{L}_{cGAN}(G, D) &= \mathbb{E}_{\mathbf{X}_{mD}^{noisy}, \mathbf{X}_{mD}^{clean}} \left[ \operatorname{mean}\left(\log \mathbf{M}_{out}^{real}\right) \right] \\
		& + \mathbb{E}_{\mathbf{X}_{mD}^{noisy}} \left[ \operatorname{mean}\left(\log (\mathbf{1} - \mathbf{M}_{out}^{fake}) \right) \right].
	\end{split}
\end{equation}

An $L_1$ distance penalty loss is also imposed to ensure the generated spectrum is structurally close to the ground truth, as
\begin{equation}
    \mathcal{L}_{L1}(G) = \mathbb{E}_{\mathbf{X}_{mD}^{noisy}, \mathbf{X}_{mD}^{clean}} \left[ \| \mathbf{X}_{mD}^{clean} - \hat{\mathbf{X}}_{mD} \|_1 \right].
\end{equation}
The final optimization problem is a minimax game given by\begin{equation}G^* = \arg \min_G \max_D (\mathcal{L}_{cGAN}(G, D) + \lambda_{loss} \mathcal{L}_{L1}(G)),\end{equation}
where $\lambda_{loss}$ is the balancing parameter to control the weight of the reconstruction error.

It can be observed from Fig.~\ref{cGAN} that the noisy mD spectrum is severely corrupted by noise, which results in obscured micro-motion patterns. 
In contrast, the denoised mD spectrum exhibits a significantly clear micro-motion pattern. Hence, the cGAN-based denoising network successfully suppresses the noise while preserving the structural integrity of the mD signatures.
\begin{figure*}[!t]
	\centering
	\includegraphics[width=180mm]{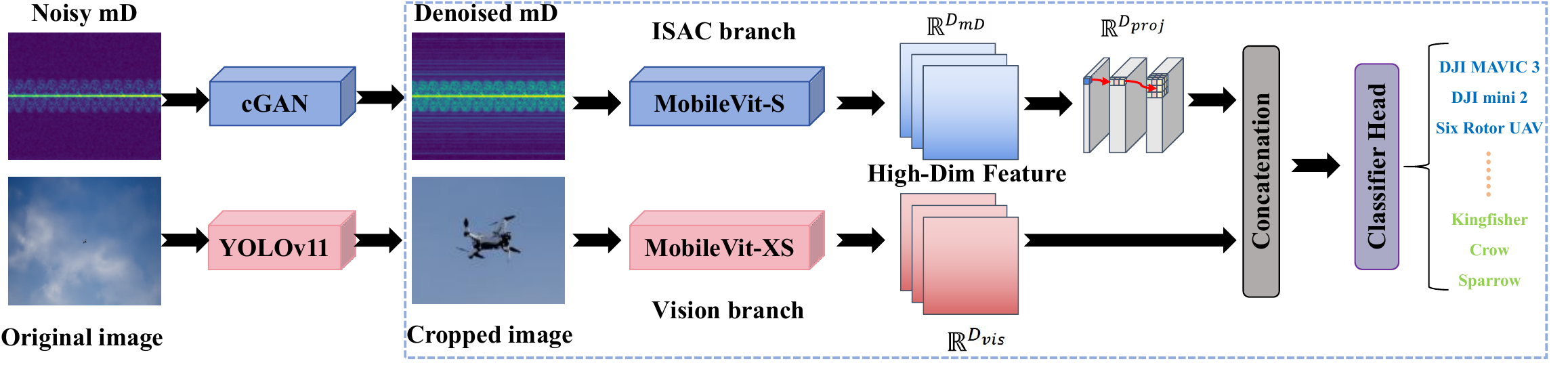}
	\caption{Fine-grained low-altitude target recognition network with MobileViT.}
	\label{model}
\end{figure*}
\section{Fine-Grained Low-Altitude Target Recognition Network with MobileViT}

In this section,
we utilize YOLOv11 to preprocess the visual image, 
and then design a fine-grained low-altitude target recognition network with MobileViT.

\subsection{Visual ROI Extraction by YOLOv11}
We employ an ROI extraction module to eliminate background redundancy, which directs the subsequent recognition network to only focus on the target's features. Specifically, we utilize the YOLOv11 detector  to localize the low-altitude targets within the RGB images. Let $\mathbf{X}_{vis}^{raw}$ denote the original image. The detector outputs the bounding box coordinates $\mathcal{B} = [x_c, y_c, w_b, h_b]$ of the target, where $(x_c, y_c)$ represents the center coordinates, \textcolor{black}{while $w_b, h_b$ denote the width and height of the box, respectively.} \textcolor{black}{
To extract the target, we convert the bounding box into pixel coordinates which correspond to the top-left $(u_1, v_1)$ and bottom-right $(u_2, v_2)$ corners, i.e.,}
\begin{align}
	u_1 &= x_c - w_b/2, \quad v_1 = y_c - h_b/2, \\
	u_2 &= x_c + w_b/2, \quad v_2 = y_c + h_b/2.
\end{align}
The small-sized feature image $\mathbf{X}_{ROI}$ is cropped from the original image based on the indices as
\begin{equation}
\mathbf{X}_{ROI} = \mathbf{X}_{vis}^{raw}(v_1:v_2, u_1:u_2, :).
\end{equation} 
Then we normalize  $\mathbf{X}_{ROI}$ to a standardized dimension $\mathbf{X}_{vis} \in \mathbb{R}^{H_{v} \times W_{v} \times 3}$ by bilinear interpolation, which aligns with the input specifications of the following recognition network.

\subsection{Fine-Grained Low-Altitude Target Recognition Network}

Since the visual images and mD spectrum possess distinct statistical properties, we design a dual-stream architecture to extract latent features individually. 
As shown in Fig.~\ref{model}, we adopt MobileViT as the backbone for both ISAC and vision branches. MobileViT effectively combines the spatial inductive biases of convolutional neural network (CNN) with the global processing capabilities of ViT, which makes it highly suitable to capture both the fine-grained features of low-altitude targets in RGB images and the periodic micro-motion patterns in mD spectrum.

The vision branch processes the pre-processed RGB tensor $\mathbf{X}_{vis}$. We utilize the MobileViT-XS variant to maintain a lightweight deployment profile. Let $\mathcal{F}_{vis}(\cdot)$ denote the feature mapping function instantiated by the MobileViT-XS backbone. The feature mapping function progressively downsamples the input $\mathbf{X}_{vis}$ while aggregating global context via self-attention mechanisms. To obtain a compact representation, we apply global average pooling (GAP) to the final feature maps, which yields the feature vector $\mathbf{f}_{vis} \in \mathbb{R}^{D_{vis}}$ as
\begin{equation}
\mathbf{f}_{vis} = \text{GAP}(\mathcal{F}_{vis}(\mathbf{X}_{vis})),
\end{equation}
where $D_{vis}$ represents the channel dimension of the extracted visual features.

The ISAC branch takes the aligned mD spectrum $\mathbf{X}_{mD}$ as the input. We employ a slightly deeper model, MobileViT-S, to capture the complex time-frequency couplings of micro-motions.
Specifically, the convolutional layers capture local spectral textures, such as sharp Doppler shifts. The transformer layers model the long-range temporal correlations inherent in periodic micro-motion signatures that often span across multiple time frames.
Let $\mathcal{F}_{mD}(\cdot)$ denote the mapping function of the ISAC backbone. The extracted feature vector $\mathbf{f}_{mD}^{raw} \in \mathbb{R}^{D_{mD}}$ is obtained as
\begin{equation}
\mathbf{f}_{mD}^{raw} = \text{GAP}(\mathcal{F}_{mD}(\mathbf{X}_{mD})),
\end{equation}
where $D_{mD}$ corresponds to the output dimension of MobileViT-S.

To enhance the representation capability of the mD branch, we map the raw sensing feature $\mathbf{f}_{mD}^{raw}$ to a higher-dimensional space, which can be expressed as
\begin{equation}
\mathbf{f}_{mD} = \sigma_{ReLU} \left( \text{BN} \left( \mathbf{W}_{proj} \mathbf{f}_{mD}^{raw} + \mathbf{b}_{proj} \right) \right),
\end{equation}
where $\mathbf{W}_{proj} \in \mathbb{R}^{D_{proj} \times D_{mD}}$ and $\mathbf{b}_{proj}$ are learnable parameters. The projection step ensures that the ISAC mD features are explicitly emphasized before fusion.

Subsequently, we concatenate visual feature vector $\mathbf{f}_{vis}$ and the projected ISAC mD  feature vector $\mathbf{f}_{mD}$ along the channel dimension to form the unified multi-modal representation as
$\mathbf{f}_{fused} = [\mathbf{f}_{vis}, \mathbf{f}_{mD}] \in \mathbb{R}^{D_{total}}, $where $D_{total} = D_{vis} + D_{proj}$. 
The fused feature is then fed into a multi-layer perceptron (MLP) classifier to predict the fine-grained target category. The network is trained end-to-end by minimizing the cross-entropy loss between the output and the ground truth labels.

    


\begin{figure}[!t]
	\centering
	\includegraphics[width=90mm]{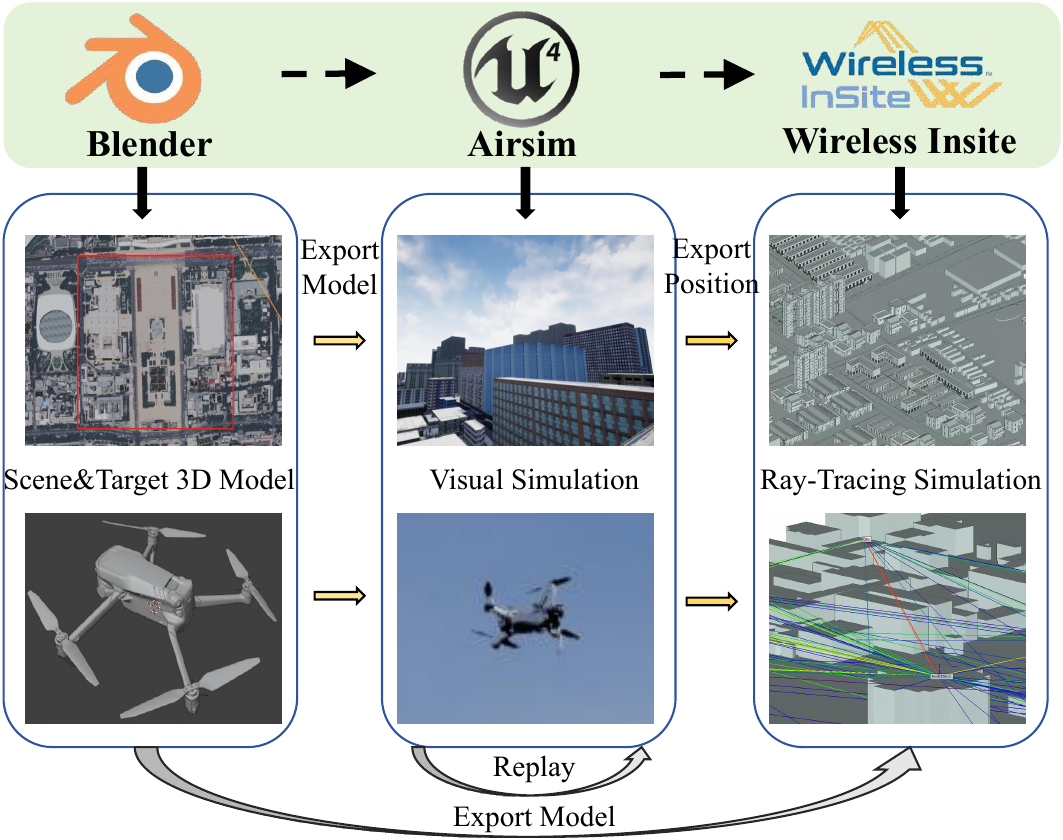}
	\caption{Generation pipeline of joint ISAC and vision low-altitude target monitoring dataset.}
	\label{UE}
\end{figure}
\begin{figure*}[!t]
	\centering
	\includegraphics[width=180mm]{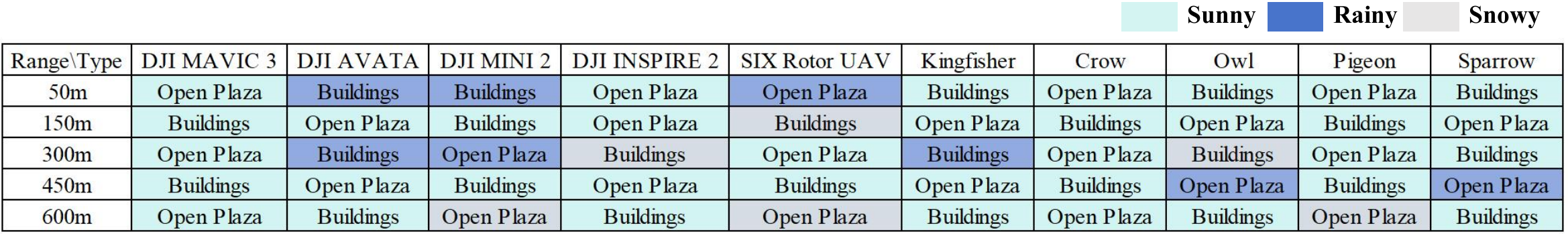}
	\caption{Composition of the genetated joint ISAC and vision low-alititude targets monitoring dataset.}
	\label{dataset}
\end{figure*}

\section{Simulation Results}

In this section, \textcolor{black}{we generate a  low-altitude targets monitoring dataset named JIVD for joint ISAC and vision recognition}, and then evaluate the performance of the proposed fine-grained
low-altitude target  recognition framework.

\subsection{RGB Image and Channel State Generation via AirSim and Wireless InSite}
As shown in Fig.~\ref{UE}, we conduct the RGB image generation of low-altitude targets on Microsoft AirSim, a high-fidelity simulator built on Unreal Engine 4 (UE4) \cite{shah2017airsim}. We import 3D mesh models of low-altitude targets and large-scale map assets from Blender into the AirSim \cite{blender}. However, \textcolor{black}{a critical challenge in JIVD generation is to ensure the strict temporal alignment between the echo signals and the RGB images}. Hence, we utilize a trajectory-replay image generation strategy.
We first control the target to execute flight behaviours within the designated monitoring zone $\mathcal{S}$. Meanwhile, we record the ground truth position and velocity at a high sampling frequency by AirSim's sensor interface.
Then we generate the RGB image frame by frame, which uses a replay approach instead of continuous recording. Specifically,
AirSim iterates through the recorded positions. For each timestamp, we pause the simulation clock and reposition the target to exact recorded position. Meanwhile, we actively configure the PTZ camera based on the target's current location, which follows the control logic (27) and (30). Once the configuration is settled, an RGB image is captured and stored.

Next, we import the 3D mesh models in Blender and positions recorded in the AirSim into Wireless InSite\cite{Remcom_WirelessInSite}. The target's position is configured by the recorded location frame by frame. 
It ensures that for every RGB image captured in the PTZ camera, there exists a corresponding echo channel state generated in the Wireless InSite.
We then utilize X3D ray-tracing solver to extract the echo channel parameters. For each frame of the target's trajectory, the solver calculates the propagation characteristics of the rays, which include the line-of-sight (LoS) path and  non-line-of-sight (NLoS) components.
Given the small physical size and relatively long detection distance of low-altitude targets, we focus exclusively on the LoS path and neglect NLoS components in this paper.
The simulation output provides a comprehensive set of geometric channel parameters for each propagation path, which includes time of arrival, angles of departure and angles of arrival. 
Finally, we substitute geometric parameters into the theoretical channel model derived in Section II to synthesize the raw echo signal data by MATLAB.

\subsection{{Composition and Settings of Generated Dataset }}

\begin{figure*}[!t]
	\centering
	\subfloat[]{\includegraphics[width=60mm]{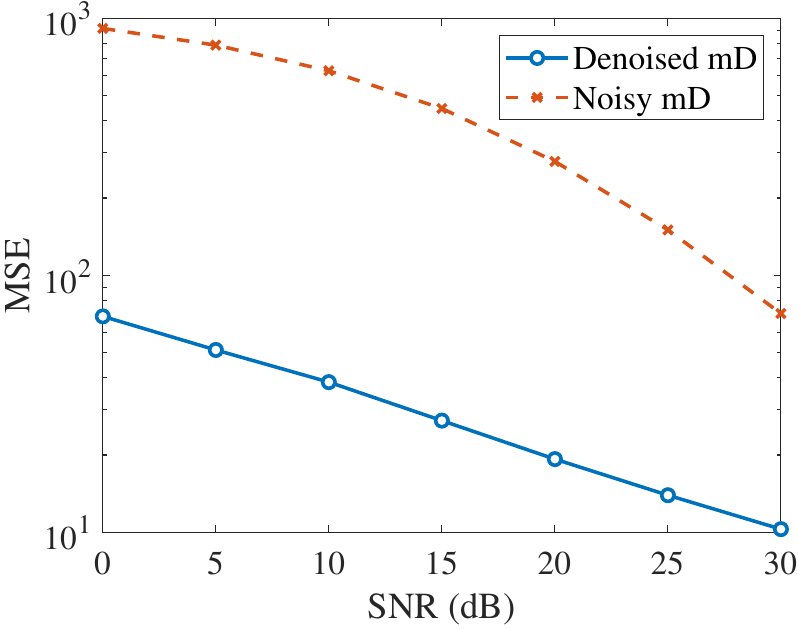}%
		\label{fig_first_case}}
	\hfil
	\subfloat[]{\includegraphics[width=60mm]{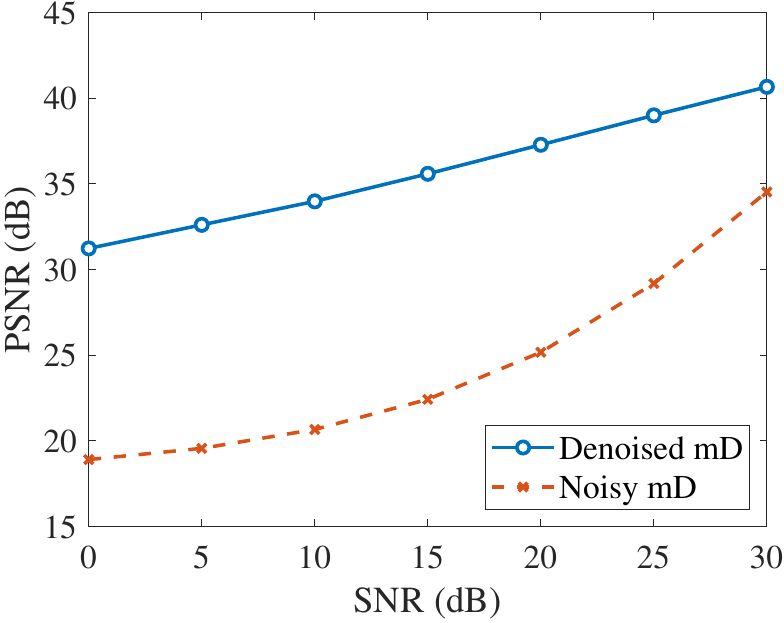}%
		\label{fig_first_case}}
	\hfil
	\subfloat[]{\includegraphics[width=60mm]{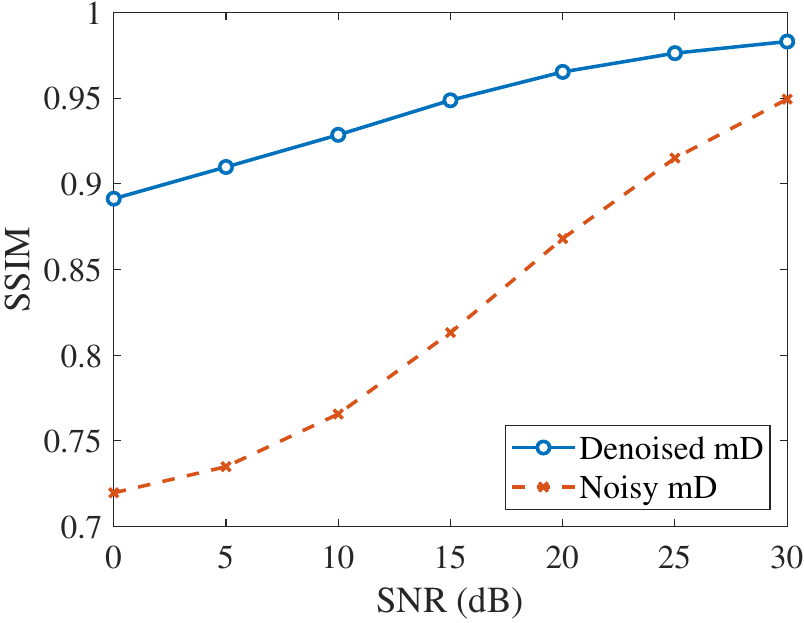}%
		\label{fig_first_case}}
	\caption{\textcolor{black}{Comparison of reconstruction quality metrics between  noisy and denoised mD spectrum under different SNR. (a) MSE. (b) PSNR. (c) SSIM.}}
	\label{psnr}
\end{figure*}

Fig.~\ref{dataset} illustrates the detailed composition of the generated dataset. The dataset includes 10 distinct subcategories of low-altitude targets for fine-grained recognition, which consists of 5 UAV models and 5 bird species. The UAVs include DJI MAVIC 3, DJI AVATA, DJI MINI~2, DJI INSPIRE 2 and  Six Rotor UAV. The birds include kingfisher, crow, owl, pigeon and sparrow.
Specific details on the modeling process, kinematics, micro-motion parameters, and geometry of these targets can be found in \cite{11159257}.
The simulation scenarios mainly cover open plazas and dense urban building complexes. Moreover, sunny, rainy and snowy weather conditions are included in AirSim. 

Furthermore, each trajectory sequence created by AirSim has a duration of 5 seconds with a sampling rate of 24 frames per second (FPS). 
To ensure diversity of the dataset, targets are generated with initial position ranges set at 50 m, 150 m, 300 m, 450 m, and 600 m for every target class. 
Consequently, the dataset comprises a total of 10$\times$5$\times$5$\times$24 = 6,000 RGB images, each with a resolution of $1920\times1080$ pixels.
Hence, the corresponding parameter $L_{cls}$ for fine-grained target recognition is set at 40 pixels.
Subsequently, the positions generated by AirSim are imported into Wireless InSite to perform channel state generation. The results are then exported to MATLAB to generate 6,000 frames of echo signals synchronized with the visual images.

\begin{figure}[!t]
	\centering
	\includegraphics[width=80mm]{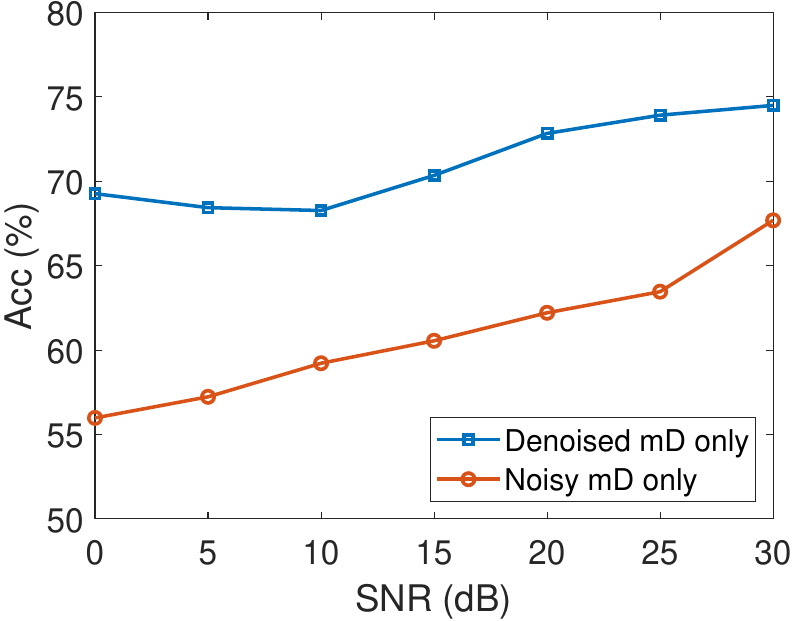}
	\caption{Testing recognition accuracy of noisy mD and denoised mD.}
	\label{2acc}
\end{figure}

We set the operating frequency of  BS as $f_0 = 26$~GHz, the subcarrier spacing as $\Delta f =  100$ kHz,
the number of subcarriers as $M = 1024$, the antenna spacing as $d_x = d_z = d = \frac{\lambda}{2} = \frac{c}{2f_0}$, the array size of HU-URA as $N^{x}_{H}\times N_{H}^z = 16 \times 16$, the array size of RU-URA as $N^{x}_{R}\times N_{R}^z = 8 \times 8$, the number of OFDM symbols contained in one frame as $N = 4200$, and the OFDM symbol interval as $T_s = 10~\mu\text{s}$. 
For time-frequency analysis,  we set the window length for STFT as $L_{win} = 
128$ with a step size $A_{step} = 4$.
Then we add random noise with  SNR belonging to \{0, 5, 10, 15, 20, 25, 30\}~dB to the noiseless echo signal, which results in  $6,000\times7 = 42,000$ noisy echo signals.
Upon generation, the dataset is partitioned into training and testing sets.
For sequences shown  in Fig.~\ref{dataset}, we randomly choose $80\%$ sequences for training  and $20\%$ sequences for testing.

\begin{figure}[!t]
	\centering
	\includegraphics[width=80mm]{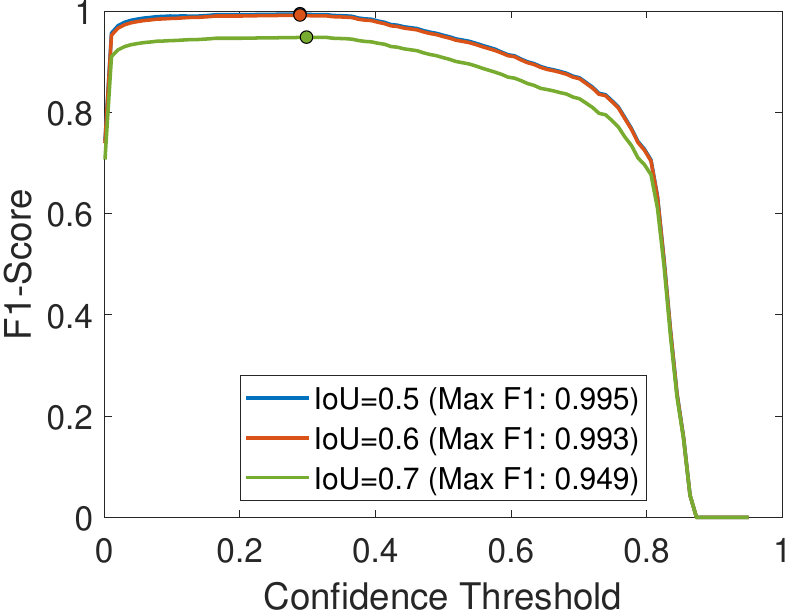}
	\caption{F1-score performance under varying confidence threshold at IoU thresholds of 0.5, 0.6, and 0.7.}
	\label{F1}
\end{figure}
\subsection{{Performance of  mD Spectrum Denoising Enhancement}}
During the  cGAN-based denoising network training phase,
we pair each clean spectrum
$\mathbf{X}_{mD}^{clean}$ with its corresponding noisy spectrum $\mathbf{X}_{mD}^{noisy}$ across all SNR levels as input. 
In the testing phase, only the unseen noisy $\mathbf{X}_{mD}^{noisy}$ from the testing set is input. 
To evaluate the performance of the proposed cGAN-based denoising network, we introduce three standard metrics in image reconstruction: mean squared error (MSE), peak signal-to-noise ratio (PSNR), and structural similarity index measure (SSIM) \cite{1284395}, respectively expressed as
\begin{align}
	\text{MSE} &= \frac{1}{H_r W_r} \big\| \mathbf{X}_{mD}^{clean} - \hat{\mathbf{X}}_{mD} \big\|_F^2, \\
	\text{PSNR} &= 10 \log_{10} \left( \frac{\max(\mathbf{X}_{mD}^{clean})^2}{\text{MSE}} \right), \\
	\text{SSIM} &= \frac{(2\mu_c \mu_r + C_1)(2\sigma_{cr} + C_2)}{(\mu_c^2 + \mu_r^2 + C_1)(\sigma_c^2 + \sigma_r^2 + C_2)},
\end{align}
where $\| \cdot \|_F$ denotes the Frobenius norm, $\mu_c$ and $\mu_r$ denote the mean intensities of $\mathbf{X}_{mD}^{clean}$ and $\hat{\mathbf{X}}_{mD}$, $\sigma_c^2$ and $\sigma_r^2$ are the corresponding variances, $\sigma_{cr}$ is the covariance,  and $C_1, C_2$ are constants included to ensure numerical stability.

\begin{figure}[!t]
	\centering
	\includegraphics[width=80mm]{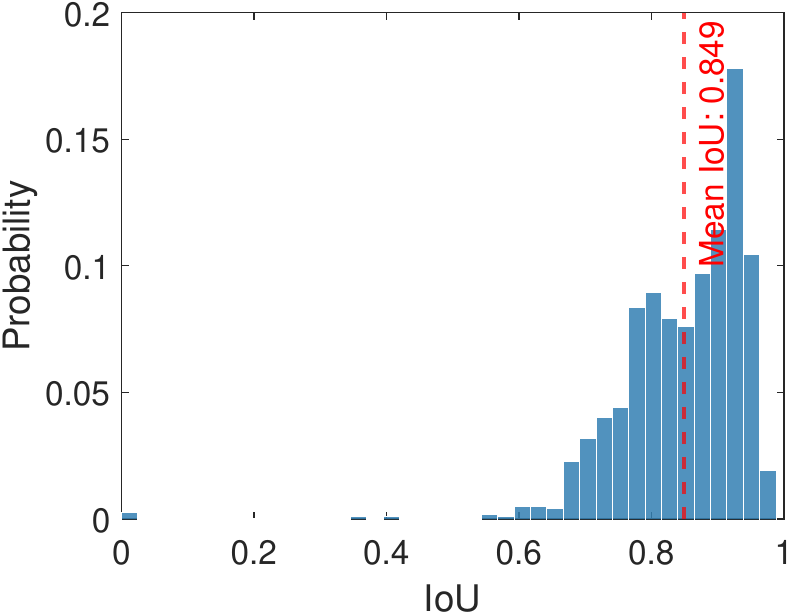}
	\caption{Histogram of IoU on test dataset}
	\label{IOU}
\end{figure}

Fig.~\ref{psnr} presents the  comparison between the noisy mD spectrum and the denoised mD spectrum. It can be observed from Fig.~\ref{psnr}(a), Fig.~\ref{psnr}(b), and Fig.~\ref{psnr}(c) that the denoised mD spectrum consistently outperforms the original ones. 
\textcolor{black}{Specifically, the denoised mD spectrum possesses significantly higher PSNR and SSIM, while possessing lower MSE, which indicates that the proposed  cGAN-based denoising network effectively suppresses the noise while preserving the structural integrity of the target's micro-motion signatures.} 

Furthermore, to verify the contribution of the cGAN-based denoising network to the recognition performance, we evaluate the recognition accuracy of the ISAC only branch with and without denoising. For clarity, we denote these two schemes as ``Denoised mD only'' and ``Noisy mD only'', respectively.
Fig.~\ref{2acc} illustrates the accuracy curves versus SNR. As shown in the figure, the performance of the  ``Noisy mD only'' scheme degrades rapidly as the SNR decreases. In contrast, the ``Denoised mD only'' scheme maintains a relatively higher accuracy, which shows a significant performance gap.

\begin{figure}[!t]
	\centering
	\includegraphics[width=80mm]{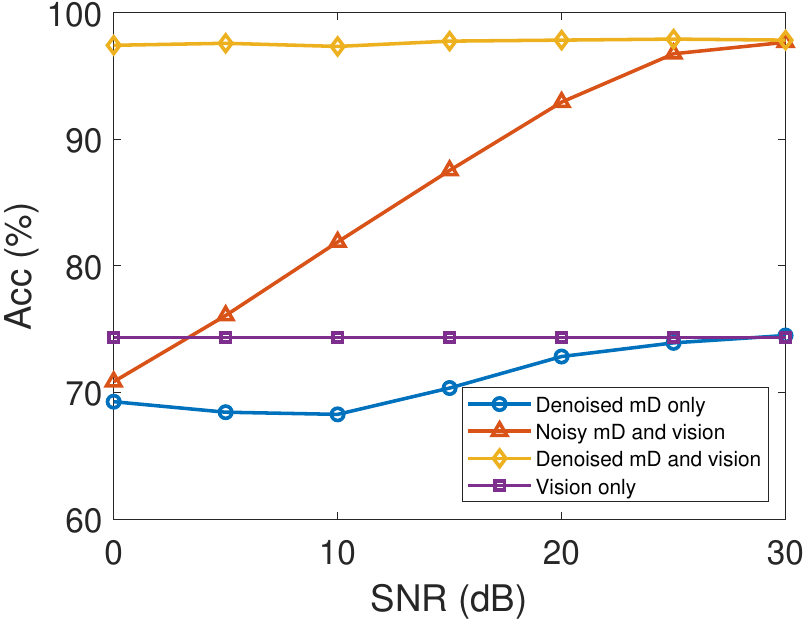}
	\caption{Testing recognition accuracy versus SNRs of different schemes}
	\label{4acc}
\end{figure}
\begin{figure}[!t]
	\centering
	\includegraphics[width=80mm]{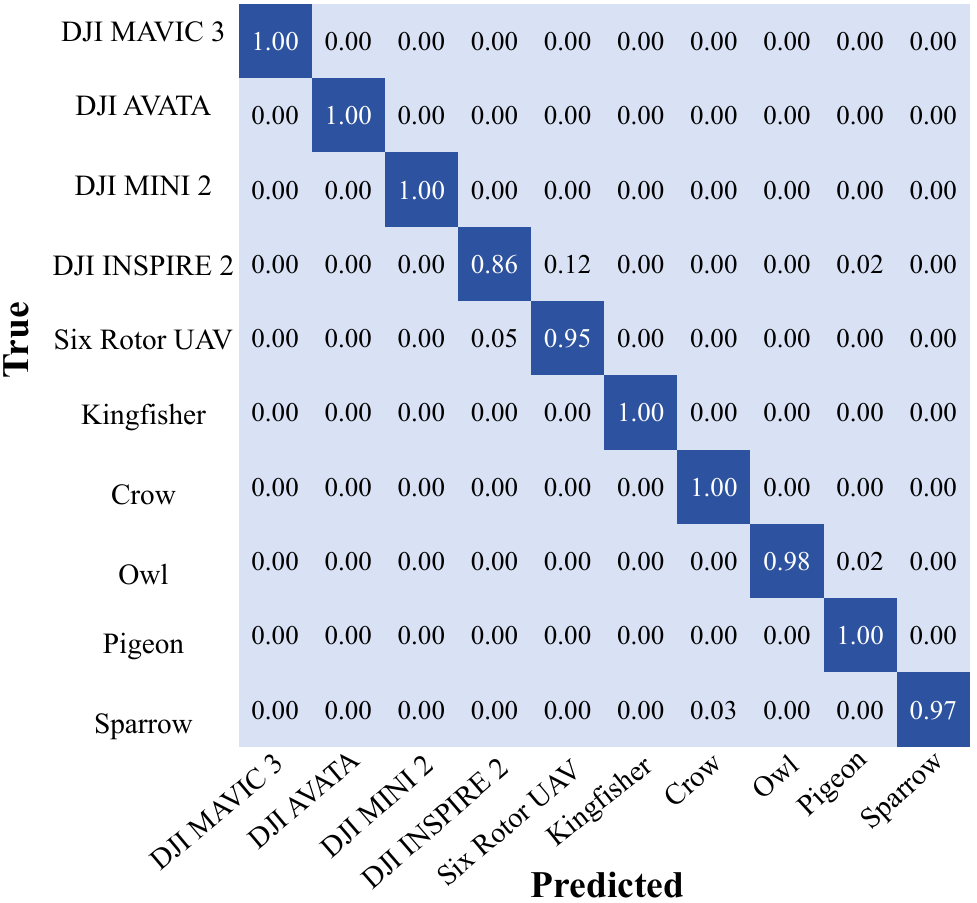}
	\caption{Confusion matrices of 30 dB test data}
	\label{cm}
\end{figure}
\subsection{Performance of Visual ROI Extraction by YOLOv11}

Since the vision branch relies on cropped small-sized feature images for feature extraction, the localization quality of YOLOv11 is important for the overall framework performance.
Hence, we use class-agnostic metrics to focus on YOLOv11's  ability that could separate low-altitude targets from the background. We introduce the following metrics based on the intersection over union (IoU) threshold. Let $N_{TP}$ denote the number of true positives, defined as detections satisfying the condition $\text{IoU} \ge \text{threshold}$. Conversely, $N_{FP}$ represents the number of false positives where $\text{IoU} < \text{threshold}$. Moreover, $N_{FN}$ indicates the number of false negatives, which corresponds to the targets missed by the detector.
The precision ($P$) and recall ($R$) are formulated as\begin{equation}P = \frac{N_{TP}}{N_{TP} + N_{FP}}, \quad R = \frac{N_{TP}}{N_{TP} + N_{FN}}.\end{equation}
Then the F1-score is calculated as\begin{equation}F_1 = 2 \cdot \frac{P \cdot R}{P + R}.\end{equation}

Fig.~\ref{F1} depicts the F1-score curves under 0.5, 0.6, and 0.7 IoU thresholds. It can be observed that YOLOv11 maintains a high F1-score across a wide range of confidence thresholds. Moreover, the F1-score achieves value of 0.995 at the  threshold of $\text{IoU}=0.5$, which implies that there rarely exists false alarms and missed detections within the RGB images.

The histogram of IoU is illustrated in Fig.~\ref{IOU}. The majority of samples fall within the high-IoU range with a mean IoU of 0.849, which confirms that the YOLOv11 achieves precise localization capabilities for low-altitude targets. The high-precision cropping guarantees that the vision branch receives consistently aligned visual features, which provides a reliable foundation for the ISAC and vision fusion.

\subsection{{\textcolor{black}{Performance of Fine-Grained Low-Altitude Target Recognition Network}}}

\begin{figure*}[!t]
	\centering
	\includegraphics[width=174mm]{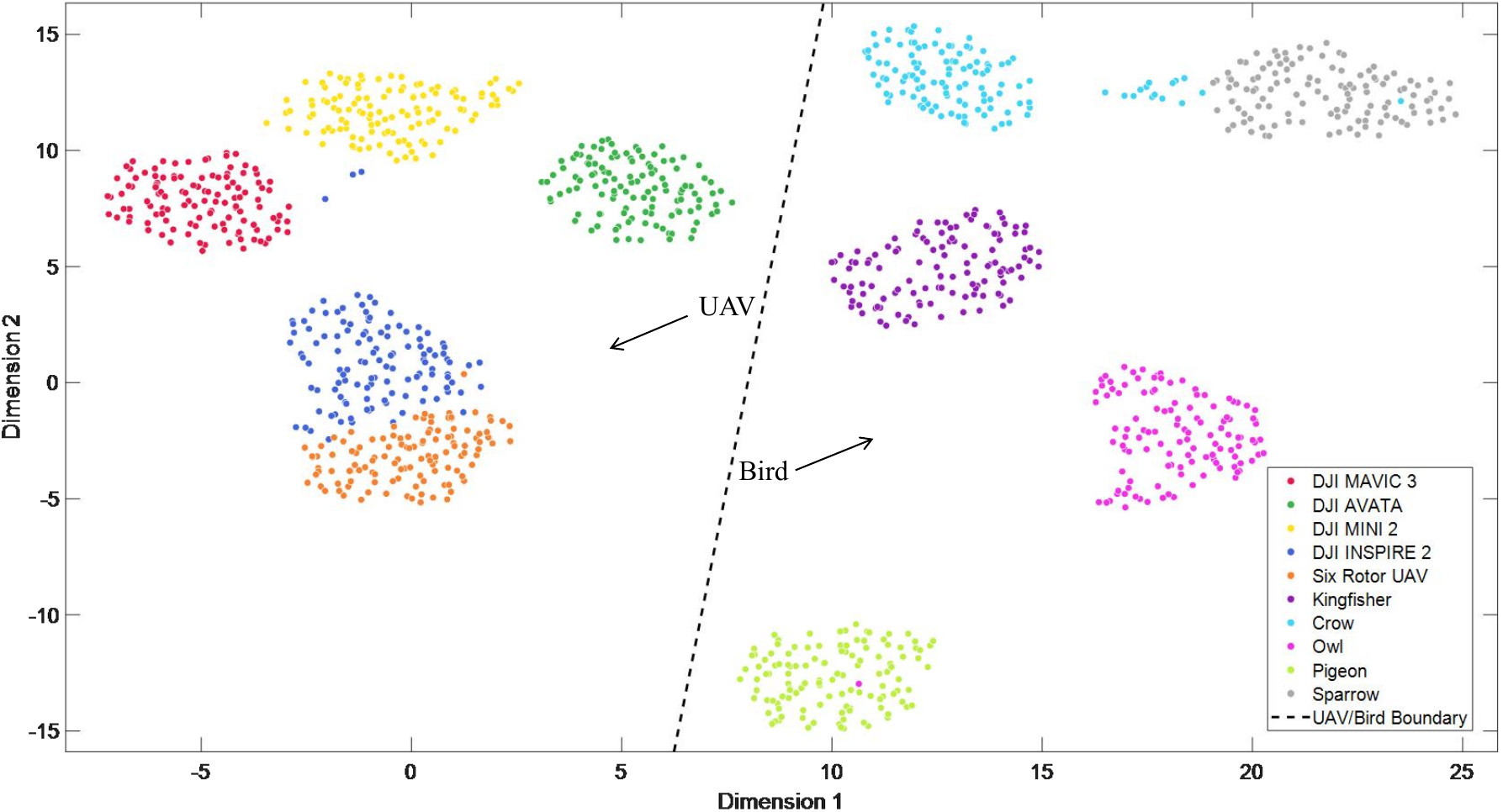}
	\caption{UMAP analysis of 30 dB test data}
	\label{UMAP}
\end{figure*}

Fig.~\ref{4acc} compares the fine-grained recognition accuracy of four experimental settings under varying SNRs.
For clarity, we denote the scheme that fuses denoised mD spectrum and RGB images as ``Denoised mD and vision'', while the fusion of noisy mD spectrum and RGB images is termed as ``Noisy mD and vision''. Additionally, the scheme that only uses RGB images is referred to as ``Vision only''.
It is evident that the proposed ``Denoised mD and vision'' scheme achieves the highest accuracy and maintains remarkable stability across the entire SNR range with an average accuracy of $97.67\%$.
In contrast, the ``Noisy mD and vision'' scheme degrades significantly at low SNRs with an average accuracy of $86.25\%$, which requires a high SNR of 30 dB to eventually match the performance of the proposed method.
Meanwhile, the single-modal schemes exhibit significantly inferior performance. The average accuracy of the ``Vision only'' scheme is limited to $74.34\%$, while the ``Denoised mD only'' scheme achieves only $71.08\%$. The substantial performance gap verifies the necessity of dual-modal fusion. The integration of features provides critical complementary information that a single sensor cannot offer.
In Fig.~\ref{cm}, we illustrate the confusion matrix of the ``Denoised mD and vision'' scheme for the 10-category recognition. As observed from the figure, the diagonal elements are clearly dominant, which indicates high recognition accuracy across all specific classes.

To further validate the recognition performance of the proposed dual-stream fusion network, we utilize the uniform manifold approximation and projection (UMAP) algorithm to visualize the high-dimensional feature distribution \cite{mcinnes2018umap}. 
Specifically, we extract the fused feature vectors $\mathbf{f}_{fused}$ from the layer before MLP. Then the $\mathbf{f}_{fused}$ are projected onto a 2D plane for visualization. 
Fig.~\ref{UMAP} illustrates the feature UMAP, where each point represents a test sample, color-coded by its ground truth category.
It can be observed from Fig.~\ref{UMAP} that the samples corresponding to the 10 target categories form distinct, compact clusters. The visualization demonstrates high inter-class separability and low intra-class variance, which indicates that the network has successfully learned robust and exclusive feature representations for each fine-grained target.

To investigate the semantic distinction between the low-altitude targets, we perform a binary classification analysis on the projected 2D manifold. A linear support vector machine (SVM) is employed to determine the decision boundary between the UAV super-class and the bird super-class \cite{708428}. As depicted by the dashed line in Fig.~\ref{UMAP}, a clear linear boundary exists that perfectly separates the UAV clusters from the bird clusters. The margin between the two super-classes verifies that the fused features are highly discriminative, which provides a solid basis for reliable target type identification.


\section{Conclusions}


In this paper, we have proposed a fine-grained low-altitude target recognition framework based on ISAC and vision fusion system.
We first estimated the position of the  low-altitude target based on ISAC system.
Then, we derived the working parameters of the PTZ camera based on the estimated  target position, such that the camera could capture the image of tiny low-altitude target from several hundred meters away.
Subsequently, we utilized STFT to obtain the mD spectrum of the target, and designed  a cGAN-based denoising network to optimize the quality of the mD spectrum.
Meanwhile, we employed YOLOv11 to detect the low-altitude target and  crop small-sized feature image.
Next, we  designed a fine-grained low-altitude target recognition network with MobileViT to distinguish the subcategory of low-altitude target.
Moreover, we generated JIVD based on AirSim and Wireless InSite, which includes diverse target categories, scenarios, and weather conditions.
The effectiveness and superiority of the  proposed    scheme   have been demonstrated by simulation results.

\bibliographystyle{ieeetr}
\bibliography{AAAref.bib}

\begin{thebibliography}{10}
\providecommand{\url}[1]{#1}
\csname url@samestyle\endcsname
\providecommand{\newblock}{\relax}
\providecommand{\bibinfo}[2]{#2}
\providecommand{\BIBentrySTDinterwordspacing}{\spaceskip=0pt\relax}
\providecommand{\BIBentryALTinterwordstretchfactor}{4}
\providecommand{\BIBentryALTinterwordspacing}{\spaceskip=\fontdimen2\font plus
\BIBentryALTinterwordstretchfactor\fontdimen3\font minus
  \fontdimen4\font\relax}
\providecommand{\BIBforeignlanguage}[2]{{%
\expandafter\ifx\csname l@#1\endcsname\relax
\typeout{** WARNING: IEEEtran.bst: No hyphenation pattern has been}%
\typeout{** loaded for the language `#1'. Using the pattern for}%
\typeout{** the default language instead.}%
\else
\language=\csname l@#1\endcsname
\fi
#2}}
\providecommand{\BIBdecl}{\relax}
\BIBdecl

\bibitem{Jiang20236GNN}
Y.~{Jiang} \emph{et~al.}, ``{6G} non-terrestrial networks enabled low-altitude
  economy: Opportunities and challenges,'' \emph{arXiv e-prints}, p.
  arXiv:2311.09047, Nov. 2023.

\bibitem{10723207}
H.~Huang, J.~Su, and F.-Y. Wang, ``The potential of low-altitude airspace: The
  future of urban air transportation,'' \emph{IEEE Trans. Intell. Veh.},
  vol.~9, no.~8, pp. 5250--5254, Aug. 2024.

\bibitem{9275613}
M.~Giordani and M.~Zorzi, ``Non-terrestrial networks in the {6G} era:
  Challenges and opportunities,'' \emph{IEEE Network}, vol.~35, no.~2, pp.
  244--251, Mar. 2021.

\bibitem{9666755}
M.~A. Hoque \emph{et~al.}, ``{IoTaaS:} drone-based internet of things as a
  service framework for smart cities,'' \emph{IEEE Internet Things J.}, vol.~9,
  no.~14, pp. 12\,425--12\,439, Jul. 2022.

\bibitem{10977743}
C.~Zhao, Y.~Feng, H.~Luo, F.~Gao, F.~Liu, and S.~Jin, ``Networked {ISAC}-based
  {UAV} tracking and handover toward low-altitude economy,'' \emph{IEEE Trans.
  Wireless Commun.}, vol.~24, no.~9, pp. 7670--7685, Apr. 2025.

\bibitem{11159257}
H.~Luo \emph{et~al.}, ``{AirGuard}: {UAV} and bird recognition scheme for
  integrated sensing and communications system,'' \emph{IEEE J. Sel. Areas
  Commun.}, vol.~44, pp. 835--848, 2026.

\bibitem{9681624}
M.~Mozaffari, X.~Lin, and S.~Hayes, ``Toward {6G} with connected sky: {UAVs}
  and beyond,'' \emph{IEEE Commun. Mag.}, vol.~59, no.~12, pp. 74--80, Jan.
  2022.

\bibitem{10535988}
A.~N. Sayed \emph{et~al.}, ``Enhanced {UAV} detection and classification using
  machine learning and {MIMO} radars,'' \emph{IEEE Trans. on Microw. Theory and
  Tech.}, vol.~72, no.~11, pp. 6716--6727, Nov. 2024.

\bibitem{10569843}
B.~Ma \emph{et~al.}, ``A method of photoelectric capture of {UAV} under rough
  guidance,'' in \emph{Proc. Int. Conf. Cloud Comput. Big Data Anal.}, Chengdu,
  China, Apr. 2024, pp. 184--187.

\bibitem{doi:10.1177/1729881420962907}
S.~Chen \emph{et~al.}, ``Low-altitude protection technology of {anti-UAVs}
  based on multisource detection information fusion,'' \emph{Int. J. Adv.
  Robot. Syst.}, vol.~17, no.~5, p. 1729881420962907, 2020.

\bibitem{8353365}
Z.~Kaleem and M.~H. Rehmani, ``Amateur drone monitoring: State-of-the-art
  architectures, key enabling technologies, and future research directions,''
  \emph{IEEE Wireless Commun.}, vol.~25, no.~2, pp. 150--159, May 2018.

\bibitem{11106829}
S.~Li \emph{et~al.}, ``Domain-adaptive {UAV} recognition using {IR-UWB} radar
  and a lightweight {Mamba}-based network,'' \emph{IEEE Sensors J.}, vol.~25,
  no.~18, pp. 34\,913--34\,926, Aug. 2025.

\bibitem{8239598}
B.-S. Oh \emph{et~al.}, ``Micro-{Doppler} mini-{UAV} classification using
  empirical-mode decomposition features,'' \emph{IEEE Geosci. and Remote Sens.
  Lett.}, vol.~15, no.~2, pp. 227--231, Dec. 2018.

\bibitem{7508914}
B.~Torvik, K.~E. Olsen, and H.~Griffiths, ``Classification of birds and {UAVs}
  based on radar polarimetry,'' \emph{IEEE Geosci. and Remote Sens. Lett.},
  vol.~13, no.~9, pp. 1305--1309, Jul. 2016.

\bibitem{khanam2024yolov5}
R.~Khanam and M.~Hussain, ``What is {YOLOv5}: A deep look into the internal
  features of the popular object detector,'' \emph{arXiv e-prints}, p.
  arXiv:2407.20892, Jul. 2024.

\bibitem{11083683}
B.~Wang, J.~Li, M.~Zhou, and Q.~Lu, ``Drone detection and tracking: An
  edge-deployable efficient algorithm based on vision sensor,'' \emph{IEEE
  Sensors J.}, vol.~25, no.~17, pp. 34\,126--34\,140, Jul. 2025.

\bibitem{sohan2024review}
M.~Sohan, T.~Sai~Ram, and C.~V. Rami~Reddy, ``A review on {YOLOv8} and its
  advancements,'' in \emph{Proc. Int. Conf. Data Intell. Cogn. Inform.}, 2024,
  pp. 529--545.

\bibitem{11213400}
D.~Dosi, C.~E. Abraham, and D.~Kumar, ``Experiments on vision-based methods to
  detect, identify and track {UAVs} using thermal camera,'' in \emph{Proc. Int.
  Conf. Emerg. Technol. Auton. Aer. Veh.}, Bangalore, India, Aug. 2025, pp.
  1--6.

\bibitem{9349620}
F.~Mahdavi and R.~Rajabi, ``Drone detection using convolutional neural
  networks,'' in \emph{Proc. Iranian Conf. Signal Process. Intell. Syst.},
  Mashhad, Iran, Dec. 2020, pp. 1--5.

\bibitem{8846214}
B.~Taha and A.~Shoufan, ``Machine learning-based drone detection and
  classification: State-of-the-art in research,'' \emph{IEEE Access}, vol.~7,
  pp. 138\,669--138\,682, Sep. 2019.

\bibitem{10311246}
V.~Mehta \emph{et~al.}, ``Real-time {UAV} and payload detection and
  classification system using radar and camera sensor fusion,'' in \emph{Proc.
  IEEE/AIAA Digit. Avion. Syst. Conf.}, Barcelona, Spain, 2023, pp. 1--6.

\bibitem{11257244}
V.~Mehta \emph{et~al.}, ``Edge {AI}-enabled radar and camera integration for
  real-time drone detection and classification,'' in \emph{Proc. IEEE/AIAA
  Digit. Avion. Syst. Conf.}, Montreal, Canada, 2025, pp. 1--7.

\bibitem{Tang_2024}
H.~Tang \emph{et~al.}, ``Radar-optical fusion detection of {UAV} based on
  improved {YOLOv7-tiny},'' \emph{Meas. Sci. Technol.}, vol.~35, no.~8, p.
  085110, May 2024.

\bibitem{9296833}
A.~Zhang \emph{et~al.}, ``Perceptive mobile networks: Cellular networks with
  radio vision via joint communication and radar sensing,'' \emph{IEEE Veh.
  Technol. Mag.}, vol.~16, no.~2, pp. 20--30, Jun. 2021.

\bibitem{9737357}
F.~Liu, Y.~Cui, C.~Masouros, J.~Xu, T.~X. Han, Y.~C. Eldar, and S.~Buzzi,
  ``Integrated sensing and communications: Toward dual-functional wireless
  networks for 6{G} and beyond,'' \emph{IEEE J. Sel. Areas Commun.}, vol.~40,
  no.~6, pp. 1728--1767, Jun. 2022.

\bibitem{9040264}
M.~Giordani, M.~Polese, M.~Mezzavilla, S.~Rangan, and M.~Zorzi, ``Toward 6{G}
  networks: Use cases and technologies,'' \emph{IEEE Commun. Mag.}, vol.~58,
  no.~3, pp. 55--61, Mar. 2020.

\bibitem{11250835}
H.~Luo, T.~Zhang, C.~Zhao, Y.~Wang, B.~Lin, Y.~Jiang, D.~Luo, and F.~Gao,
  ``Integrated sensing and communications framework for {6G} networks,''
  \emph{IEEE Wireless Commun.}, vol.~32, no.~6, pp. 102--109, Dec. 2025.

\bibitem{9330512}
C.~De~Lima \emph{et~al.}, ``Convergent communication, sensing and localization
  in {6G} systems: An overview of technologies, opportunities and challenges,''
  \emph{IEEE Access}, vol.~9, pp. 26\,902--26\,925, Jan. 2021.

\bibitem{9681870}
C.~Chaccour, M.~N. Soorki, W.~Saad, M.~Bennis, P.~Popovski, and M.~Debbah,
  ``Seven defining features of terahertz {(THz)} wireless systems: A fellowship
  of communication and sensing,'' \emph{IEEE Commun. Surv. Tut.}, vol.~24,
  no.~2, pp. 967--993, Jan. 2022.

\bibitem{9144301}
M.~Z. Chowdhury, M.~Shahjalal, S.~Ahmed, and Y.~M. Jang, ``{6G} wireless
  communication systems: Applications, requirements, technologies, challenges,
  and research directions,'' \emph{IEEE Open J. Commun. Soc.}, vol.~1, pp.
  957--975, Jul. 2020.

\bibitem{10906066}
J.~Tang, Y.~Yu, C.~Pan, H.~Ren, D.~Wang, J.~Wang, and X.~You, ``Cooperative
  {ISAC}-empowered low-altitude economy,'' \emph{IEEE Trans. Wireless Commun.},
  vol.~24, no.~5, pp. 3837--3853, May 2025.

\bibitem{11151696}
Y.~Huang, J.~Yang, S.~Xia, C.-K. Wen, and S.~Jin, ``Learned off-grid imager for
  low-altitude economy with cooperative {ISAC} network,'' \emph{IEEE Trans.
  Wireless Commun.}, vol.~25, pp. 3333--3348, 2026.

\bibitem{11077832}
J.~Wei, D.~Ma, F.~He, Q.~Zhang, Z.~Feng, Z.~Liu, and T.~Liang, ``{UAV’s}
  rotor micro-doppler feature extraction using integrated sensing and
  communication signal: Algorithm design and testbed evaluation,'' \emph{IEEE
  Trans. Wireless Commun.}, vol.~24, no.~12, pp. 10\,166--10\,182, Jul. 2025.

\bibitem{11069481}
J.~Xue, Q.~Zhang, D.~Ma, and J.~Wei, ``{DC-Former} network empowered {UAV} and
  bird recognition based on integrated sensing and communication system,'' in
  \emph{Proc. Int. Conf. Comput. Commun. Syst.}, Chengdu, China, Apr. 2025,
  pp.927-932.

\bibitem{4570206}
J.~Chen, Y.-c. Wu, S.~Ma, and T.-s. Ng, ``{Ml} joint {CFO} and channel
  estimation in {OFDM} systems with timing ambiguity,'' \emph{IEEE Trans.
  Wireless Commun.}, vol.~7, no.~7, pp. 2436--2440, Jul. 2008.

\bibitem{goodfellow2014generative}
I.~J. Goodfellow \emph{et~al.}, ``Generative adversarial nets,'' in \emph{Adv.
  Neural Inf. Process. Syst.}, vol.~27, 2014.

\bibitem{khanam2024yolov11}
R.~Khanam and M.~Hussain, ``{Yolov11}: An overview of the key architectural
  enhancements,'' \emph{arXiv e-prints}, p. arXiv:2410.17725, Oct. 2024.

\bibitem{mehta2021mobilevit}
S.~Mehta and M.~Rastegari, ``{MobileViT}: Light-weight, general-purpose, and
  mobile-friendly vision transformer,'' \emph{arXiv e-prints}, p.
  arXiv:2110.02178, Oct. 2021.

\bibitem{7523373}
R.~Shafin, L.~Liu, J.~Zhang, and Y.-C. Wu, ``{DoA} estimation and capacity
  analysis for {3-D} millimeter wave massive-{MIMO/FD-MIMO} {OFDM} systems,''
  \emph{IEEE Trans. Wireless Commun.}, vol.~15, no.~10, pp. 6963--6978, Oct.
  2016.

\bibitem{9947033}
Z.~Du, F.~Liu, W.~Yuan, C.~Masouros, Z.~Zhang, S.~Xia, and G.~Caire,
  ``Integrated sensing and communications for {V2I} networks: Dynamic
  predictive beamforming for extended vehicle targets,'' \emph{IEEE Trans.
  Wireless Commun.}, vol.~22, no.~6, pp. 3612--3627, Jun. 2023.

\bibitem{10634315}
C.~Du and Y.~Jiang, ``Broad beam designs for broadcast channels,'' \emph{IEEE
  Trans. Signal Process.}, vol.~72, pp. 3819--3833, Aug. 2024.

\bibitem{ronneberger2015u}
O.~Ronneberger, P.~Fischer, and T.~Brox, ``U-net: Convolutional networks for
  biomedical image segmentation,'' in \emph{Proc. Int. Conf. Med. Image Comput.
  Comput.-Assist. Interv.}, 2015, pp. 234--241.

\bibitem{8100115}
P.~Isola, J.-Y. Zhu, T.~Zhou, and A.~A. Efros, ``Image-to-image translation
  with conditional adversarial networks,'' in \emph{Proc. IEEE Conf. Comput.
  Vis. Pattern Recognit.}, 2017, pp. 5967--5976.

\bibitem{shah2017airsim}
S.~Shah, D.~Dey, C.~Lovett, and A.~Kapoor, ``{AirSim}: High-fidelity visual and
  physical simulation for autonomous vehicles,'' in \emph{Proc. Int. Conf.
  Field Serv. Robot.}, 2017, pp. 621--635.

\bibitem{blender}
\BIBentryALTinterwordspacing
B.~O. Community, \emph{Blender - a 3D modelling and rendering package}, Blender
  Foundation, Stichting Blender Foundation, Amsterdam, 2018. [Online].
  Available: \url{http://www.blender.org}
\BIBentrySTDinterwordspacing

\bibitem{Remcom_WirelessInSite}
\BIBentryALTinterwordspacing
Remcom, ``Wireless insite.'' [Online]. Available:
  \url{https://www.remcom.com/wireless-insite-em-propagation-software}
\BIBentrySTDinterwordspacing

\bibitem{1284395}
Z.~Wang, A.~Bovik, H.~Sheikh, and E.~Simoncelli, ``Image quality assessment:
  from error visibility to structural similarity,'' \emph{IEEE Trans. Image
  Process.}, vol.~13, no.~4, pp. 600--612, Apr. 2004.

\bibitem{mcinnes2018umap}
L.~McInnes, J.~Healy, and J.~Melville, ``Umap: Uniform manifold approximation
  and projection for dimension reduction,'' \emph{arXiv e-prints}, p.
  arXiv:1802.03426, Sep. 2018.

\bibitem{708428}
M.~Hearst, S.~Dumais, E.~Osuna, J.~Platt, and B.~Scholkopf, ``Support vector
  machines,'' \emph{IEEE Intell. Syst. Their Appl.}, vol.~13, no.~4, pp.
  18--28, Aug. 1998.

\end{thebibliography}

\vfill

\end{document}